\documentclass[fleqn,usenatbib]{mnras}

\usepackage[T1]{fontenc}
\usepackage{ae,aecompl}

\usepackage{graphicx}	
\usepackage{amsmath}	
\usepackage{color}

\hypersetup{final}

\usepackage{mathptmx}
\usepackage{txfonts}

\newcommand{\gizmo}{{\sc Gizmo}}

\newcommand{\anarchy}{{\sc Anarchy}}
\newcommand{\swift}{{\sc Swift}}
\newcommand{\eagle}{{EAGLE}}

\newcommand{\phantomsph}{{PHANTOM}}
\newcommand{\sphenix}{{\sc Sphenix}}

\title[\sphenix{}]{\sphenix{}: Smoothed Particle Hydrodynamics for the next generation of galaxy formation simulations}
\author[Borrow et al.]{
Josh Borrow$^{1, 2}$,
Matthieu Schaller$^{3,4}$,
Richard G. Bower$^{1}$, and
Joop Schaye$^{4}$
\\$^1$Institute for Computational Cosmology, Department of Physics, University of Durham, South Road, Durham, DH1 3LE, UK
\\$^2$Department of Physics, Kavli Institute for Astrophysics and Space Research, Massachusetts Institute of Technology, Cambridge, MA 02139, USA
\\$^3$Lorentz Institute for Theoretical Physics, Leiden University, PO Box 9506, NL-2300 RA Leiden, the Netherlands
\\$^4$Leiden Observatory, Leiden University, P.O. Box 9513, 2300 RA Leiden, The Netherlands
}

\begin{document}

\maketitle

\begin{abstract}
    Smoothed Particle Hydrodynamics (SPH) is a ubiquitous numerical method for
solving the fluid equations, and is prized for its conservation properties,
natural adaptivity, and simplicity. We introduce the \sphenix{} SPH scheme,
which was designed with three key goals in mind: to work well with sub-grid
physics modules that inject energy, be highly computationally efficient (both
in terms of compute and memory), and to be Lagrangian. \sphenix{} uses a
Density-Energy equation of motion, along with variable artificial viscosity
and conduction, including limiters designed to work with common sub-grid
models of galaxy formation. In particular, we present and test a novel
limiter that prevents conduction across shocks, preventing spurious radiative
losses in feedback events. \sphenix{} is shown to solve many difficult test
problems for traditional SPH, including fluid mixing and vorticity
conservation, and it is shown to produce convergent behaviour in all tests
where this is appropriate. Crucially, we use the same parameters within
\sphenix{} for the various switches throughout, to demonstrate the performance
of the scheme as it would be used in production simulations. \sphenix{} is the
new default scheme in the \swift{} cosmological simulation code and is available
open-source.
\end{abstract}


\section{Introduction}

There have been many approaches to solving the equations of motion for a
collisional fluid in a cosmological context over the years, from simple first
order fixed-grid \citep{Cen1992} to high-order discontinuous Galerkin schemes
solved in an adaptive environment \citep{Guillet2019}. Because Smoothed
Particle Hydrodynamics (SPH) strikes the sweet spot between computational
cost, stability, and adaptivity, it has been used throughout the astronomical
community for nearly five decades.

SPH was originally developed by \citet{Lucy1977} and \citet{Gingold1977a}, and
was used initially to model individual stars, as this problem was well suited
to Lagrangian schemes. Shortly after, further applications of the method were
deployed for the study of the fragmentation of gas clouds \citep{Wood1981},
and for the formation of planets \citep{Benz1988}.

The practical use of SPH in a cosmological context began with
\citet{Hernquist1989}, which provided a novel solution to the large dynamic
range of time-steps required to evolve a cosmological fluid, and was cemented
by the Gadget-2 code \citep{Springel2005} that was made public and exploited
worldwide to model galaxy formation processes within this context for the
first time \citep[e.g.][]{Dolag2004, Ettori2006, Crain2007}. The base SPH
model released in Gadget-2, however, was relatively simple, consisting of a
fixed artificial viscosity coefficient and scheme based on
\citet{Monaghan1992}. Improved models existed, such as those presented in
\citet{Monaghan1997}, but the key that led to the community rallying around
Gadget-2 was both its open source nature and scalability, with Gadget-2 able
to run on hundreds or thousands of cores.

The popularity of Gadget-2, and similar codes like GASOLINE
\citep{Wadsley2004}, along with its relatively simple hydrodynamics model,
led to critical works such as \citet{Agertz2007} and \citet{Bauer2012} that
pointed out flaws in their SPH modelling, relative to mesh-based codes of the
time. The SPH community as a whole, however, already had solutions to these
problems \citep[see e.g.][]{Price2008} and many robust solutions were
proposed and integrated into cosmological modelling codes. In
\citet{Hess2010}, the authors experimented with an extension to Gadget-2
using a Voronoi mesh to reduce errors inherrent in SPH and allow for better
results on fluid mixing problems, eventually giving rise to the AREPO moving
mesh scheme, allowing for significantly improved accuracy per particle but
drastically increasing computational cost \citep{Springel2010,
Weinberger2020}. In this case, the authors have steadily increased their
computational cost per particle in an attempt to reduce errors inherrent in
their hydrodynamics model as much as practicable.

Other authors took different directions, with the GASOLINE code
\citep{Wadsley2004, Wadsley2008, Wadsley2017} choosing to explicitly average
pressures within the SPH equation of motion to alleviate the problems of
artificial surface tension; the PHANTOM developers \citep{Price2008, Price2012,
Price2018} advocating for artificial conduction of energy; and further
developments on the Gadget-2 and updated Gadget-3 code by \citet{Hopkins2013}
and \citet{Hu2014} based on the work by \citet{Saitoh2013} using an explicit
smoothed pressure scheme to ensure a consistent pressure field over the contact
discontinuities that artificial surface tension arises from.

Simultaneously, there was work to reduce the fundamental numerical errors
present in SPH taking place by \citep{Cullen2010, Dehnen2012, Read2010a,
Read2012} through the use of improved choices for the SPH kernel, which up
until this point was assumed to have little effect on results from SPH
simulations. These improved kernels typically have larger `wings',
encompassing more neighbours and providing more accurate reconstructions for
smoothed quantities. These more accurate reconstructions are particularly
important for the construction of accurate gradients, which enter into
`switches' that control the strength of the artificial viscosity and
conduction terms.

The rise of more complex SPH models occurred alongside a significant jump in
the complexity of the corresponding galaxy formation models; such an increase
in complexity was required as resolutions increased over time, meaning more
physics could be modelled directly. Many astrophysical processes take place
on scales smaller than what can be resolved in simulations and are included
in these so-called galaxy formation `sub-grid' models. These processes
include radiative cooling, which has progressed from a simple one parameter
model to element and even ionisation state dependent rates \citep[see e.g.
][]{Wiersma2009, Ploeckinger2020}; star formation \citep[see e.g.][and
references therein]{Cen1992a, Schaye2008}; and stellar feedback to model
supernovae and element outflows \citep[see e.g. ][and references
therein]{Navarro1993,Springel2003, DallaVecchia2008,DallaVecchia2012}. The
coupling of these processes to hydrodynamics is complex and often overlooked;
careful treatment of conservation laws and quirks of the chosen variables
used to represent the fluid can frequently hide errors in plain sight
\citep{Borrow2020b}.

The development of the \swift{} code \citep{Schaller2016} led to a
re-implementation of the sub-grid model used for the \eagle{} simulation
\citep{Schaye2015}, and a chance to re-consider the \anarchy{} SPH scheme that
was used in the original (Gadget-3 based) code \citep[see][for details on the
scheme]{Schaller2015}. The findings in \citet{Oppenheimer2018} (their
Appendix D) and \citet{Borrow2020b} meant that a switch away from the original
Pressure-Entropy scheme to one based on a smoothed density field was preferred,
along with the key design goals outlined below. This work describes the
\sphenix{}\footnote{Note that, similar to the popular \gizmo{} schemes,
\sphenix{} is not an acronym.} scheme and demonstrates its performance on many
hydrodynamics tests. We note here that \sphenix{} does not give the best
performance-per-particle (in both absolute values of L1 norm (see \S
\ref{sec:sodshock} for our definition of the L1 norm) compared to the analytical
solution, and in terms of convergence speed) compared to other schemes. The
moving mesh AREPO \citep{Springel2010}, finite-volume GIZMO \citep{Hopkins2015},
and corrected scheme presented in \citet{Rosswog2020a} will produce improved
results. \sphenix{} however lies in the very low-cost (memory and computation)
per particle sweet-spot that traditional SPH schemes occupy, whilst maximising
performance with some novel limiters for artificial conduction and viscosity.
This makes it an excellent choice for the default SPH scheme in \swift{}.

The remainder of this paper is organised as follows: in \S \ref{sec:swift} we
describe the \swift{} cosmological simulation code and the time-stepping
algorithms present within it. In \S \ref{sec:sphenix} we describe \sphenix{} in
its entirety. In \S \ref{sec:conductionlimiter} we describe the artificial
conduction limiter used for energetic feedback schemes. Finally, in \S
\ref{sec:hydrotests} we show how \sphenix{} performs on various hydrodynamics
problems.
\section{The \swift{} simulation code}
\label{sec:swift}

The \swift{}\footnote{ For the interested reader, the implementation of the
\sphenix{} scheme was developed fully in the open and is available in the
\swift{} repository at \url{http://swiftsim.com} \citep{Schaller2018},
including all of the tests and examples shown below. We use version 0.9.0 of
the \swift{} code for the tests in this work.} simulation
code \citep{Schaller2016, Schaller2018} is a hybrid parallel SPH and gravity
code, designed to run across multiple compute nodes using MPI, but to utilise
threads on each node (rather than the traditional method of using one MPI
rank per core). This, along with its task-based parallelism approach,
asynchronous communication scheme, and work-splitting domain decomposition
system allow for excellent strong- and weak-scaling characteristics
\citep{Borrow2018}.

\swift{} is also designed to be hugely modular, with hydrodynamics schemes,
gravity schemes, and sub-grid models able to be easily swapped out. \swift{}
can be configured to use a replica of the Gadget-2 hydrodynamics scheme
\citep{Springel2002}, a simplified version of the base PHANTOM scheme
\citep{Price2018}, the MFM and MFV schemes described in \citet{Hopkins2015},
\sphenix{}, or a host of other schemes. It can also be configured to use
multiple different galaxy formation sub-grid models, including a very basic
scheme (constant $\Lambda$ cooling, no star formation), the EAGLE sub-grid
model \citep{Schaye2015}, a `Quick Lyman-$\alpha$" model, the GEAR sub-grid
model \citep{Revaz2012}, and some further evolutions including cooling tables
from \citet{Ploeckinger2020}. The gravity solver is interchangeable but the
one used here, and throughout all \swift{} simulations, uses the Fast
Multipole Method \citep{Greengard1987} with an adaptive opening angle,
similar to \citet{Dehnen2014}.

\subsection{Time integration}

\swift{} uses a velocity-verlet scheme to integrate particles through time.
This takes their acceleration ($\vec{a}$) from the equation of motion and
time-step ($\Delta t$) and integrates their position forward in time
through a Kick-Drift-Kick scheme as follows:
\begin{align}
    \vec{v}\left(t + \frac{\Delta t}{2}\right) & =
        \vec{v}(t) + \frac{\Delta t}{2}\vec{a}(t),\\
    \vec{r}\left(t + \Delta t\right) & =
        \vec{r}(t) + \vec{v}\left(t + \frac{\Delta t}{2}\right)\Delta t, \\
    \vec{v}\left(t + \Delta t\right) & = \vec{v}
        \left(t + \frac{\Delta t}{2}\right) + \frac{\Delta t}{2}\vec{a}(t+\Delta t),
    \label{eqn:KDK}
\end{align}
where the first and last equations, updating the velocity, are referred to as
the `kick', and the central equation is known as the `drift'. The careful
observer will note that the `drift' can be split into as many pieces as
required allowing for accurate interpolation of the particle position
in-between kick steps. This is important in cosmological galaxy formation
simulations, where the dynamic range is large. In this case, particles are
evolved with their own, particle-carried time-step, given by
\begin{equation}
    \Delta t_i = {\rm C}_{\rm CFL}\frac{2 \gamma_K h_i}{v_{{\rm sig}, i}},
    \label{eqn:timestep}
\end{equation}
dependent on the Courant–Friedrichs–Lewy \citep[C$_{\rm CFL}$][]{Courant1928}
constant, the kernel-dependent relationship between cut-off and smoothing
length $\gamma_K$, particle-carried smoothing length $h_i$, and signal
velocity $v_{{\rm sig}, i}$ (see Equation \ref{eqn:sigvelindiv}). The
discussion of the full time-stepping algorithm is out of the scope of this
work, but see \citet{Hernquist1989} and \citet{Borrow2019} for more
information.

\subsubsection{Time-step Limiter}

As the time-step of the particles is particle-carried, there may be certain
parts of the domain that contain interacting particles with vastly different
time-steps (this is particularly promoted by particles with varied
temperatures within a given kernel). Having these particles interact is
problematic for a number of reasons, and as such we include the time-step
limiter described in \citet{Saitoh2009, Durier2012} in all problems solved below.
\swift{} chooses to limit neighbouring particles to have a maximal time-step
difference of a factor of 4.
\section{\sphenix{}}
\label{sec:sphenix}

The \sphenix{} scheme was designed to replace the \anarchy{} scheme
used in the original \eagle{} simulations for use in the \swift{}
simulation code. This scheme had three major design goals:
\begin{itemize}
    \item Be a Lagrangian SPH scheme, as this has many advantages and is
          compatible with the \eagle{} subgrid model.
    \item Work well with the \eagle{} subgrid physics, namely 
          instantaneous energy injection and subgrid cooling.
    \item Be highly computationally and memory efficient.
\end{itemize}
The last requirement precludes the use of any Riemann solvers in so-called
\gizmo{}-like schemes \citep[although these do not necessarily give improved
results for astrophysical problem sets, see ][]{Borrow2019}; see Appendix
\ref{app:particlecost}. The second requirement also means that the use of a
pressure-based scheme (such as \anarchy{}) is not optimal, see
\citet{Borrow2020b} for more details.

The \sphenix{} scheme is based on so-called `Traditional' Density-Energy
SPH. This means that it uses the smoothed mass density,
\begin{equation}
    \hat{\rho}(\vec{x}) = \sum_j m_j W(|\vec{x} - \vec{x}_j|, h(\vec{x}))
    \label{eqn:massdensity}
\end{equation}
where here $j$ are indices describing particles in the system, $h(\vec{x})$
is the smoothing length evaluated at position $\vec{x}$, and $W(r, h)$ is the
kernel function.

In the examples below, the Quartic Spline (M5) kernel,
\begin{equation}
    w(q) = \begin{cases}
        \left(\frac{5}{2} - q\right)^4 - 5\left(\frac{3}{2} - q\right)^4 + 10\left(\frac{1}{2} - q\right)^4 & q < \frac{1}{2} \\
        \left(\frac{5}{2} - q\right)^4 - 5\left(\frac{3}{2} - q\right)^4 & \frac{1}{2} \leq q < \frac{3}{2} \\
        \left(\frac{5}{2} - q\right)^4 &  \frac{3}{2} \leq q < \frac{5}{2}  \\
        0 & q \geq \frac{5}{2}\\
    \end{cases}
    \label{eqn:quarticspline}
\end{equation}
with $W(r, h) = \kappa_{n_D} w(r / h) / h^{n_D}$, $n_D$ the number of
dimensions, and $\kappa_3 = (7 / 478 \pi)$ for three dimensions, is used. The
\sphenix{} scheme has also been tested with other kernels, notably the Cubic
and Quintic Spline (M4, M6) and the Wendland (C2, C4, C6) kernels
\citep{Wendland1995}. The choice of kernel does not qualitatively affect the
results in any of the tests in this work \citep[see][for significantly more
information on kernels]{Dehnen2012}. Higher order kernels do allow for lower
errors on tests that rely on extremely accurate reconstructions to cancel forces
(for instance the Gresho-Chan vortex, \S \ref{sec:gresho}), but we find that the
Quintic Spline provides an excellent trade-off between computational cost and
accuracy in practical situations. Additionally, the Wendland kernels do have the
benefit that they are not susceptible to the pairing instability, but they must have
an ad-hoc correction applied in practical use \citep[][Section 2.5]{Dehnen2012}.
We find no occurrences of the pairing instability in both the tests and our
realistic simulations. The \sphenix{} scheme is kernel-invariant, and
as such can be used with any reasonable SPH kernel.

The smoothing length $h$ is determined by satisfying
\begin{equation}
    \hat{n}(\vec{x}) = \sum_j W\left(|\vec{x} - \vec{x}_j|, h(\vec{x})\right)=
        \left(\frac{\eta}{h(\vec{x})}\right)^{n_D},
    \label{eqn:findh}
\end{equation}
with $\eta$ setting the resolution scale. The precise choice for $\eta$
generally does not qualitatively change results; here we choose $\eta=1.2$ due
to this value allowing for a very low $E_0$ error \citep[see ][]{Read2010a,
Dehnen2012}\footnote{This corresponds to  $\sim$58 weighted neighbours for our
Quartic Spline in a scenario where all neighbours have uniform smoothing
lengths. In practical simulations the `number of neighbours' that a given
particle interacts with can vary by even orders of magnitude but Equation
\ref{eqn:findh} must be satisfied for all particles ensuring an accurate
reconstruction of the field. More discussion on this choice of smoothing length
can be found in \citep{Springel2002, Monaghan2002, Price2007, Price2012,
Borrow2020b}. We chose $\eta=1.2$ based on Figure 3 in \citet{Dehnen2012},
where this corresponds to a very low reconstruction error in the density.},
which is a force error originating from particle disorder within a single
kernel. In \swift{}, these equations are solved numerically to a relative
accuracy of $10^{-4}$.

The smoothed mass density, along with a particle-carried internal energy per
unit mass $u$, is used to determine the pressure at a particle position
through the equation of state
\begin{equation}
    P(\vec{x}_i) = P_i = (\gamma - 1) u_i \hat{\rho}_i,
    \label{eqn:equationofstate}
\end{equation}
with $\gamma$ the ratio of specific heats, taken to be $5/3$ throughout
unless specified. This pressure enters the first law of thermodynamics,
\begin{equation}
    \left.\frac{\partial u_i}{\partial \vec{q}_i}\right|_{A_i} = 
    - \frac{P_i}{m_i} \frac{\partial V_i}{\partial \vec{q}_i},
    \label{eqn:firstlaw}
\end{equation}
with $\vec{q}_i$ a state vector containing both $\vec{x}_i$ and $h_i$ as
\emph{independent} variables, $A_i$ the entropy of particle $i$ (i.e. this
equation only applies to dissipationless dynamics), and $V_i = m_i /
\hat{\rho}_i$ describing the volume represented by particle $i$. This constraint,
along with the one on the smoothing length, allows for an equation of motion to
be extracted from a Lagrangian \citep[see e.g. the derivations in ][]{
Springel2002, Hopkins2013},
\begin{equation}
    \frac{\mathrm{d}\vec{v}_i}{\mathrm{d}t} = - \sum_j m_j
    \left[
        \frac{f_{ij} P_i}{\hat{\rho}_i^2} \nabla_i W_{ij} +
        \frac{f_{ji} P_j}{\hat{\rho}_j^2} \nabla_j W_{ji}
    \right],
    \label{eqn:veleom}
\end{equation}
where $W_{ab} = W(|\vec{x}_b - \vec{x}_a|, h(\vec{x}_a))$, $\nabla_a =
\partial / \partial \vec{x}_a$, and $f_{ab}$ a dimensionless factor
encapsulating the non-uniformity of the smoothing length field
\begin{equation}
    f_{ab} = 1 - \frac{1}{m_b} \left(
        \frac{h_a}{n_D \hat{n}_a} \frac{\partial \hat{\rho}_i}{\partial h_i}
    \right) \left(
        1 + \frac{h_i}{n_D \hat{n}_i} \frac{\partial \hat{n}_i}{\partial h_i}
    \right)^{-1}
    \label{eqn:fij}
\end{equation}
and is generally of order unity. There is also an associated equation of motion
for internal energy,
\begin{equation}
    \frac{\mathrm{d}u_i}{\mathrm{d}t} = - \sum_j m_j f_{ij} \frac{P_i}{\hat{\rho}_i^2} 
        \vec{v}_{ij} \cdot \nabla_i W_{ij},
    \label{eqn:ueom}
\end{equation}
with $\vec{v}_{ij} = \vec{v}_j - \vec{v}_i$. Note that other differences
between vector quantities are defined in a similar way, including for
the separation of two particles $\vec{x}_{ij} = \vec{x}_j - \vec{x}_i$.

\subsection{Artificial Viscosity}

These equations, due to the constraint of constant entropy introduced in the beginning,
lead to naturally dissipationless solutions; they cannot capture shocks. Shock capturing
in SPH is generally performed using `artificial viscosity'.

The artificial viscosity implemented in \sphenix{} is a simplified and
modified extension to the \citet{Cullen2010} `inviscid SPH' scheme. This
adds the following terms to the equation of motion \citep[see ][and references
within]{Monaghan1992},
\begin{equation}
    \frac{\mathrm{d}\vec{v}_i}{\mathrm{d}t} = -
    \sum_j m_j \zeta_{ij} \left[
        f_{ij} \nabla_i W_{ij} + f_{ji} \nabla_j W_{ji}
    \right],
    \label{eqn:eomviscvel}
\end{equation}
and to the associated equation of motion for the internal energy,
\begin{equation}
    \frac{\mathrm{d}u_i}{\mathrm{d}t} = - \frac{1}{2}
    \sum_j m_j \zeta_{ij} \vec{v}_{ij} \cdot \left[
        f_{ij} \nabla_i W_{ij} + f_{ji} \nabla_j W_{ji}
    \right],
    \label{eqn:eomviscu}
\end{equation}
where $\zeta_{ij}$ controls the strength of the viscous interaction. Note here
that the internal energy equation of motion is explicitly symmetrised, which was
not the case for the SPH equation of motion for internal energy (Eqn.
\ref{eqn:ueom}). In this case, that means that there are terms from both the
$ij$ and $ji$ interactions in Equation \ref{eqn:eomviscu}, whereas in Equation
\ref{eqn:ueom} there is only a term from the $ij$ interaction.  This choice was
due to the symmetric version of this equation performing significantly better in
the test examples below, likely due to multiple time-stepping errors within
regions where the viscous interaction is the strongest\footnote{For these
reasons all recent works choose symmetric forms for these equations.}.

There are many choices available for $\zeta_{ij}$, with the case used here being
\begin{equation}
    \zeta_{ij} = - \alpha_V \mu_{ij} \frac{v_{{\rm sig},ij}}
                                                    {\hat{\rho_i} + \hat{\rho_j}},
    \label{eqn:nuij}
\end{equation}
where
\begin{equation}
    \mu_{ij} = \begin{cases}
        \frac{\vec{v}_{ij} \cdot \vec{x}_{ij}}{|\vec{x}_{ij}|} & \vec{v}_{ij} \cdot \vec{x}_{ij} < 0 \\
        0 & \vec{v}_{ij} \cdot \vec{x}_{ij} \geq 0 \\
    \end{cases}
    \label{eqn:muij}
\end{equation}
is a basic particle-by-particle converging flow limiter (meaning that
the viscosity term vanishes when $\nabla \cdot \vec{v} \geq 0$), and
\begin{equation}
    v_{{\rm sig}, ij} = c_i + c_j - \beta_V\mu_{ij},
    \label{eqn:vsigij}
\end{equation}
is the signal velocity between particles $i$ and $j$, with $\beta_V=3$ a
dimensionless constant, and with $c_i$ the soundspeed of particle $i$ defined
through the equation of state as
\begin{equation}
    c_i = \sqrt{\frac{P_i}{\hat{\rho}_i}} = \sqrt{(\gamma - 1) \gamma u_i }.
    \label{eqn:soundspeed}
\end{equation}
Finally, the dimensionless viscosity coefficient $\alpha_V$
\citep{Monaghan1983} is frequently taken to be a constant of order unity. In
\sphenix{}, this becomes an interaction-dependent constant \citep[see][for
similar schemes]{Morris1997, Cullen2010}, with $\alpha_V = \alpha_{V, ij}$,
dependent on two particle-carried $\alpha$ values as follows:
\begin{equation}
    \alpha_{V, ij} = \frac{1}{4}(\alpha_{V, i} + \alpha_{V, j})(B_i + B_j),
    \label{eqn:alphavij}
\end{equation}
where
\begin{equation}
    B_i = \frac{|\nabla \cdot \vec{v}_i|}
               {|\nabla \cdot \vec{v}_i| + |\nabla \times \vec{v}_i| + 10^{-4}c_i / h_i}
    \label{eqn:balsara}
\end{equation}
is the \citet{Balsara1989} switch for particle $i$, which allows for the deactivation
of viscosity in shear flows, where there is a high value of $\nabla \cdot \vec{v}$,
but the associated shear viscosity is unnecessary. This, in particular, affects
rotating shear flows such as galaxy disks, where the scheme used to determine
$\alpha_{V, i}$ described below will return a value close to the maximum.

The equation for $\alpha_{V, i}$ is solved independently for each particle
over the course of the simulation. Note that $\alpha_{V, i}$ is never
drifted, and is only ever updated at the `kick' steps. The source term in the
equation for $\alpha_{V, i}$, designed to activate the artificial viscosity
within shocking regions, is the shock indicator
\begin{equation}
    S_i = \begin{cases}
        -h_i^2 \max\left(\dot{\nabla} \cdot \vec{v}_i, 0\right) & \nabla \cdot \vec{v}_i \leq 0 \\
        0 & \nabla \cdot \vec{v}_i > 0 \\
    \end{cases}
    \label{eqn:shockindicator}
\end{equation}
where here the time differential of the local velocity divergence field
\begin{equation}
    \dot{\nabla} \cdot \vec{v}_i (t + \Delta t) =
        \frac{\nabla \cdot \vec{v}_i (t + \Delta t) - \nabla \cdot \vec{v}_i (t)}
             {\Delta t}
    \label{eqn:divvdt}
\end{equation}
with $\nabla \cdot \vec{v}_i$ the local velocity divergence field and $\Delta
t$ the time-step associated with particle $i$. The primary variable in the
shock indicator $S_i$ of $\dot{\nabla} \cdot \vec{v}$ is high in pre-shock
regions, with the secondary condition for the flow being converging ($\nabla
\cdot \vec{v} \leq 0$) helpful to avoid false detections as the
\citet{Balsara1989} switch is used independently from the equation that
evolves $\alpha_{V, i}$ (this choice is notably different from most other
schemes that use $B_i$ directly in the shock indicator $S_i$). This choice
allows for improved shock capturing in shearing flows (e.g. feedback events
occurring within a galaxy disk). In these cases, the \citet{Balsara1989}
switch (which is instantaneously evaluated) rapidly becomes close to $1.0$,
and the already high value of $\alpha_{V, i}$ allows for a strong viscous
reaction from the fluid. The shock indicator is then transformed into an
optimal value for the viscosity coefficient as
\begin{equation}
    \alpha_{V, {\rm loc}, i} = \alpha_{V, {\rm max}}\frac{S_i}{c_i^2 + S_i},
    \label{eqn:alphaloc}
\end{equation}
with a maximum value of $\alpha_{V, {\rm max}} = 2.0$ for $\alpha_{V, {\rm
loc}}$. The value of $\alpha_{V, i}$ is then updated as follows:
\begin{equation}
    \alpha_{V, i} = \begin{cases}
        \alpha_{V, {\rm loc}, i} & \alpha_{V, i} < \alpha_{V, {\rm loc}, i} \\
        \frac{\alpha_{V, i} + \alpha_{V, {\rm loc}, i} \frac{\Delta t}{\tau_{V,i}}}
             {1 + \frac{\Delta t}{\tau_{V,i}}} & \alpha_{V, i} > \alpha_{V, {\rm loc}, i} \\
        \end{cases}
    \label{eqn:alphavevo}
\end{equation}
where $\tau_{V, i} = \gamma_K \ell_V h_i / c_i$ with $\gamma_K$ the `kernel
gamma' a kernel dependent quantity relating the smoothing length and compact
support \citep[$\gamma_K = 2.018932$ for the quartic spline in 3D,
][]{Dehnen2012} and $\ell_V$ a constant taking a value of 0.05. The final
value of $\alpha_{V, i}$ is checked against a minimum, however the default
value of this minimum is zero and the evolution strategy used above
guarantees that $\alpha_{V, i}$ is strictly positive and that the decay is
stable regardless of time-step.

\subsection{Artificial Conduction}

Attempting to resolve sharp discontinuities in non-smoothed variables in SPH
leads to errors. This can be seen above, with strong velocity discontinuities
(shocks) not being correctly handled and requiring an extra term in the
equation of motion (artificial viscosity) to be captured. A similar issue
arises when attempting to resolve strong discontinuities in internal energy
(temperature). To resolve this, we introduce an artificial energy conduction
scheme similar to the one presented by \citet{Price2008}. This adds an extra
term to the equation of motion for internal energy,
\begin{equation}
    \frac{\mathrm{d}u_i}{\mathrm{d}t} = \sum_j m_j v_{D, ij} 
        (u_i - u_j) \hat{r}_{ij} \cdot \left(
            f_{ij} \frac{\nabla_i W_{ij}}{\hat{\rho}_i} + 
            f_{ji} \frac{\nabla_j W_{ji}}{\hat{\rho}_j}
        \right)
    \label{eqn:dudtdiff}
\end{equation}
with $\hat{r}_{ij}$ the unit vector between particles $i$ and $j$, and where
\begin{equation}
    v_{D, ij} = \frac{\alpha_{D, ij}}{2}
        \left(
            \frac{|\vec{v}_{ij} \cdot \vec{x}_{ij}|}{|\vec{x}_{ij}|} + 
            \sqrt{2\frac{|P_i - P_j|}{\hat{\rho}_j + \hat{\rho}_j}}
        \right).
    \label{eqn:condspeed}
\end{equation} This conductivity speed is the average of two commonly used
speeds, with the former velocity-dependent term taken from \citet{Price2018}
\citep[modified from][]{Wadsley2008}, and the latter pressure-dependent term
taken from \citet{Price2008}. These are usually used separately for cases that
aim to reduce entropy generation in disordered fields and contact
discontinuities respectively(where initially there is a strong discontinuity in
pressure that is removed by the artificial conduction scheme), but we combine
them here as both cases are relevant in galaxy formation simulations and use
this same velocity throughout our testing, a notable difference with other
works using conduction schemes \citep[e.g.][]{Price2018}. \citet{Price2018}
avoided pressure-based terms in simulations with self-gravity, but they use no
additional terms (e.g. our $\alpha_D$) to limit conduction in flows where it is
not required. This is additionally somewhat similar to the conduction speed
used in \anarchy{} and \citet{Hu2014}, which is a modified version of the
signal velocity (Eqn. \ref{eqn:vsigij}) with our speed replacing the sum of
sound speeds with a differenced term. Appendix \ref{app:conduction} contains an
investigation of the individual terms in the conduction velocity. The
interaction-dependent conductivity coefficient,
\begin{equation}
    \alpha_{D, ij} = \frac{P_i \alpha_{D, i} + P_j \alpha_{D, j}}
                          {P_i + P_j},
    \label{eqn:alphadab}
\end{equation}
is pressure-weighted to enable the higher pressure particle to lead the
conduction interaction, a notable departure from other thermal conduction
schemes in use today. This is critical when it comes to correctly applying
the conduction limiter during feedback events, described below. The
individual particle-carried $\alpha_{D, i}$ are ensured to only be active in
cases where there is a strong discontinuity in internal energy. This is
determined by using the following discontinuity indicator,
\begin{equation}
    K_i = \beta_D \gamma_K h_i \frac{\nabla^2 u_i}{\sqrt{u_i}},
    \label{eqn:diffK}
\end{equation}
where $\beta_D$ is a fixed dimensionless constant taking a value of 1. The
discontinuity indicator enters the time differential for the individual
conduction coefficients as a source term,
\begin{equation}
    \frac{\mathrm{d}\alpha_{D, i}}{\mathrm{d}t} = K_i +
    \frac{\alpha_{D, {\rm min}} - \alpha_{D, i}}{\tau_{D, i}},
    \label{eqn:daddt}
\end{equation}
with $\tau_{D, i} = \gamma_K h_i / v_{{\rm sig}, i}$, $\alpha_{D, {\rm
min}}=0$ the minimal allowed value of the artificial conduction coefficient,
and with the individual particle signal velocity,
\begin{equation}
    v_{{\rm sig}, i} = \max_j(v_{{\rm sig}, ij}),
    \label{eqn:sigvelindiv}
\end{equation}
controlling the decay rate. $\nabla^2 u$ is used as the indicator for a
discontinuity, as opposed to $\nabla u$, as it allows for (physical, well
represented within SPH) linear gradients in internal energy to be maintained
without activating artificial conduction. This is then integrated during
`kick' steps using
\begin{equation}
    \alpha_{D, i}(t + \Delta t) = \alpha_{D, i}(t) +
            \frac{\mathrm{d}\alpha_{D, i}}{\mathrm{d}t} \Delta t.
    \label{eqn:alphaDevo}
\end{equation}
The final stage of evolution for the individual conduction coefficients is to
limit them based on the local viscosity of the fluid. This is necessary
because thermal feedback events explicitly create extreme discontinuities
within the internal energy field that lead to shocks (see \S
\ref{sec:conductionlimiter} for the motivation leading to this choice). The
limit is applied using the maximal value of viscous alpha among the neighbours
of a given particle,
\begin{equation}
    \alpha_{V, {\rm max}, i} = \max_j ( \alpha_{V, j} ),
\end{equation}
with the limiter being applied using the maximally allowed value of the
conduction coefficient,
\begin{equation}
    \alpha_{D, {\rm max}, i} = \alpha_{D, \rm{max}}\left(
                 1 - \frac{\alpha_{V, {\rm max}, i}}
                 {\alpha_{V, \rm{max}}}\right),
    \label{eqn:condshocklimiter}
\end{equation}
with $\alpha_{D, \rm{max}} = 1$ a constant, and
\begin{equation}
    \alpha_{D, i} = \begin{cases}
        \alpha_{D, i} & \alpha_{D, i} < \alpha_{D, \rm{max}} \\
        \alpha_{D, \rm{max}} & \alpha_{D, i} > \alpha_{D, \rm{max}}.
    \end{cases}
    \label{eqn:conductionlimiter}
\end{equation}
This limiter allows for a more rapid increase in conduction coefficient, and
a higher maximum, than would usually be viable in simulations with strong
thermal feedback implementations. In \anarchy{}, another scheme employing
artificial conduction, the rate at which the artificial conduction could grow
was chosen to be significantly smaller. In \anarchy{}, $\beta_D = 0.01$,
which is 100 times smaller than the value chosen here \citep[][Appendix
A3]{Schaye2015}. This additional conduction is required to accurately capture
contact discontinuities with a Density-Energy SPH equation of motion.

\section{Motivation for the Conduction Limiter}
\label{sec:conductionlimiter}

\begin{figure}
    \centering
    \includegraphics{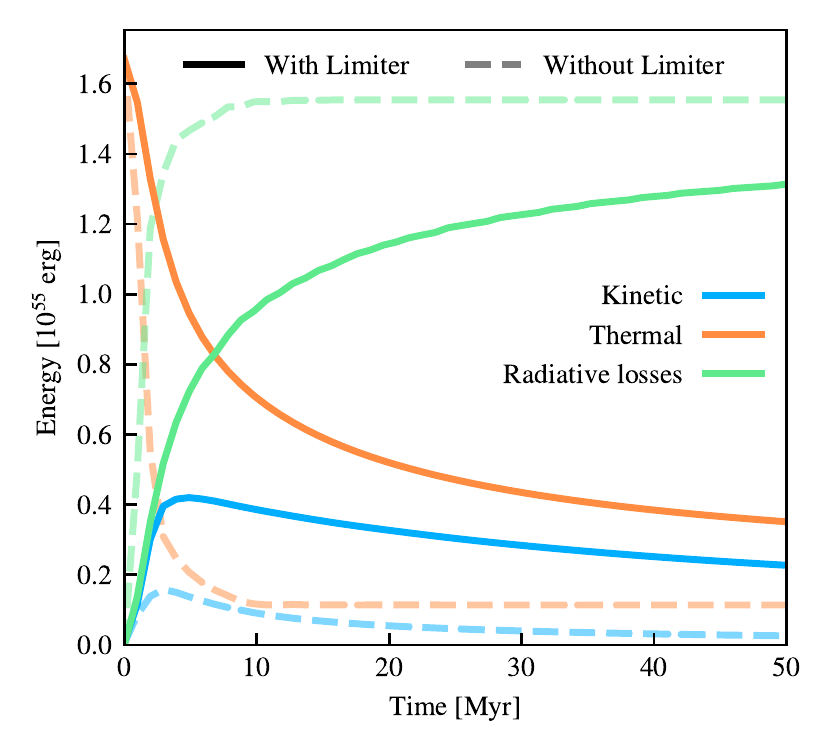}
    \vspace{-0.2cm}
    \caption{Energy in various components as a function of time for a
    simulated supernova blast (see text for details of the set-up). Blue
    shows energy in the kinetic phase, orange shows energy in the thermal
    phase (neglecting the thermal energy of the background) and green shows
    energy lost to radiation. The solid lines show the simulation performed
    with the artificial conduction limiter applied, and the dashed lines show
    the simulation without any such limiter. Simulations performed without
    the limiter show huge, rapid, cooling losses.}
    \label{fig:energycompare}
\end{figure}

\begin{figure}
    \centering
    \includegraphics{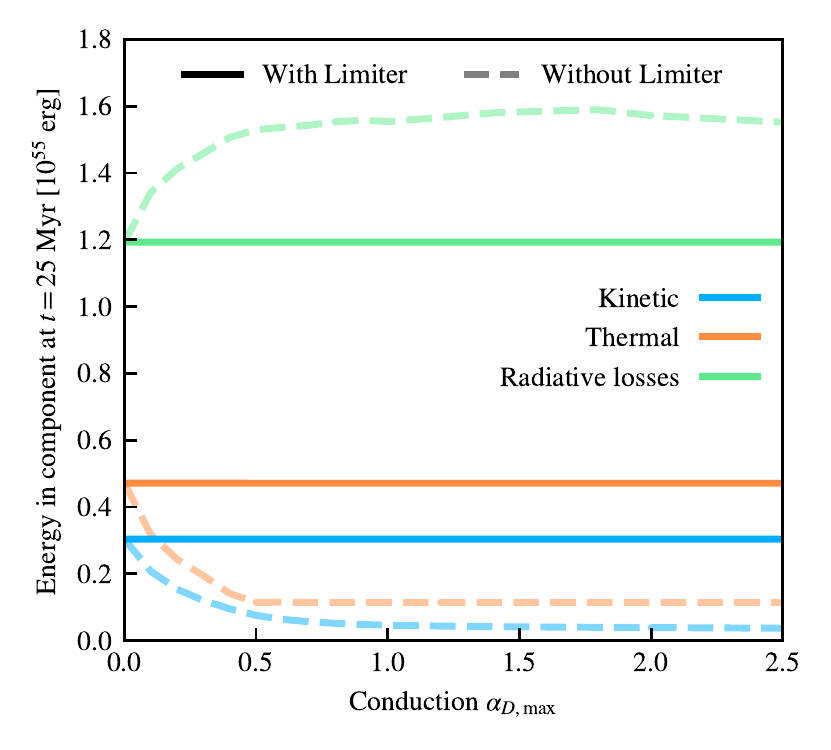}
    \vspace{-0.2cm}
    \caption{The set-up from Fig. \ref{fig:energycompare} performed for
    different values for the maximum artificial conduction coefficient
    $\alpha_{D, {\rm max}}$ (i.e. a different horizontal axis as Fig.
    \ref{fig:energycompare}, with the same vertical axis), now showing the
    components of energy in each phase at a fixed time of $t=25$ Myr. Colours
    and line styles are the same as in Fig. \ref{fig:energycompare}. As well as
    demonstrating the issue with un-limited conduction, this figure shows
    that the conduction limiter prevents the loss of additional energy energy
    relative to a simulation performed without any artificial conduction.}
    \label{fig:energyfuncdiff}
\end{figure}

The conduction limiter first described in \S \ref{sec:sphenix} is formed of
two components; a maximal value for the conduction coefficient in viscous
flows (Eqn. \ref{eqn:conductionlimiter}), and one that ensures that a
particle with a higher pressure takes preference within the conduction
interaction (Eqn. \ref{eqn:alphadab}).

This limiter is necessary due to interactions of the artificial conduction
scheme with the sub-grid physics model. Here the \eagle{} sub-grid model is
shown as this is what \sphenix{} was designed for use with, however all
schemes employing energetic feedback and unresolved cooling times will suffer
from the same problems when using un-limited artificial conduction. In short,
when an energetic feedback event takes place, the artificial conduction
switch is activated (as this is performed by injecting lots of energy into
one particle, leading to an extreme value of $\nabla^2u$). This then leads to
energy leaking out of the particle ahead of the shock front, which is then
radiated away as the neighbouring particles can rapidly cool due to their
temperature being lower leading to smaller cooling times.

To show the effect of this problem on a real system, we set up a uniform
volume containing $32^3$ gas particles at approximately solar metallicity
($Z=0.014$) and equilibrium temperature (around $10^4$ K), at a density of
$n_{\rm H} = 0.1$ cm$^{-3}$. The central particle in the volume has
approximately the same amount of energy injected into it as in a single
\eagle{}-like stellar feedback event (heating it to $\sim 10^{7.5}$ K) at the
start of the simulation and the code is ran with full sub-grid cooling
\citep[using the tables from ][]{Wiersma2009} enabled. The initial values for
the artificial viscosity and conduction coefficients are set to be zero
(whereas in practice they are set to be their maximum and minimum in `real'
feedback events; this has little effect on the results as the coefficients
rapidly stabilise).

Fig. \ref{fig:energycompare} shows the energy in the system (with the thermal
energy of the `background' particles removed to ensure a large dynamic range
in thermal energy is visible on this plot) in various components. We see
that, at least for the case with the limiter applied, at $t=0$ there is the
expected large injection of thermal energy that is rapidly partially
transformed into kinetic energy as in a classic blastwave problem (like the
one shown in Fig. \ref{fig:sedov}; in our idealised, non-radiative, Sedov
blasts only 28\% of the injected thermal energy is converted to kinetic
energy). A significant fraction, around two thirds, of the energy is lost to
radiation, but the key here is that there is a transformation of the initial
thermal injection to a kinetic wave.

In the same simulation, now with the conduction limiter removed (dashed
lines), almost all of the injected energy is immediately lost to radiation
(i.e. the feedback is unexpectedly inefficient). The internal energy in the
affected particle is rapidly conducted to its neighbours (that are then
above, but closer to, the equilibrium temperature) which have a short cooling
time and hence the energy is quickly lost.

The direct effect of the conduction limiter is shown in Fig.
\ref{fig:energyfuncdiff}, where the same problem as above is repeated ten
times with maximal artificial conduction coefficients $\alpha_{D, {\rm max}}$
of $0$ to $2.5$ in steps of $0.1$ (note that the value of $\alpha_{D, {\rm
max}}$ used in production simulations is $1$). We choose to show these
extreme values to demonstrate the efficacy of the limiter even in extreme
scenarios. The simulations with and without the limiter show the same result
at $\alpha_{D, {\rm max}} = 0$ (i.e. with conduction disabled) but those
without the limiter show a rapidly increasing fraction of the energy lost to
cooling as the maximal conduction coefficient increases. The simulations with
the limiter show a stable fraction of energy (at this fixed time of $t=25$
Myr) in each component, showing that the limiter is working as expected and
is curtailing these numerical radiative losses. This result is qualitatively
unchanged for a factor of 100 higher, or lower, density background gas (i.e.
gas between $n_{\rm H} = 0.001$ cm$^{-3}$ and $n_{\rm H} = 10.0$ cm$^{-3}$).
In both of these cases, the conduction can rapidly cause numerical radiative
losses, but with the limiter enabled this is remedied entirely. We also note
that the limiter remains effective even for extreme values of the conduction
parameter (e.g. with $\alpha_{D, {\rm max}}=100$), returning the same result
as the case without artificial conduction for this test.
\section{Hydrodynamics Tests}
\label{sec:hydrotests}

In this section the performance of \sphenix{} is shown on hydrodynamics
tests, including the \citet{Sod1978} shock tube, \citet{Sedov1959} blastwave,
and the \citet{Gresho1990} vortex, along with many other problems relevant to
galaxy formation. All problems are performed in hydrodynamics-only mode,
with no radiative cooling or any other additional physics, and
all use a $\gamma=5/3$ equation of state ($P = (2/3) u_i \hat{\rho}$).

Crucially, all tests were performed with the same scheme parameters and
settings, meaning that all of the switches are consistent (even between
self-gravitating and pure hydrodynamical tests) unless otherwise stated. This
departs from most studies where parameters are set for each problem
independently, in an attempt to demonstrate the maximal performance of the
scheme for a given test. The parameters used are as follows:
\begin{itemize}
    \item The quartic spline kernel.
    \item CFL condition $C_{\rm CFL} = 0.2$, with multiple time-stepping
          enabled \citep[see e.g.][]{Lattanzio1986}.
    \item Viscosity alpha $0.0 \leq \alpha_V \leq 2.0$ with the initial
          value being $\alpha_V = 0.1$ \citep[similar to][]{Cullen2010}.
    \item Viscosity beta $\beta_V = 3.0$ and length $\ell_V = 0.05$
    \citep[similarly to][]{Cullen2010}.
    \item Conduction alpha $0.0 \leq \alpha_D \leq 1.0$ \citep[a choice
    consistent with][]{Price2008} with the viscosity-based conduction limiter
    enabled and the same functional form for the conduction speed (Eqn.
    \ref{eqn:condspeed}) used in all simulations.
    \item Conduction beta $\beta_D = 1.0$ with the initial value of $\alpha_D
          = 0.0$.
\end{itemize}
These choices were all `calibrated' to achieve an acceptable result on the
Sod shock tube, and then left fixed with the results from the rest of the
tests left unseen. We choose to present the tests in this manner in an effort
to show a representative overview of the performance of \sphenix{} in
real-world conditions as it is primarily designed for practical use within
future galaxy formation simulations.

The source code required to produce the initial conditions (or a link to
download the initial conditions themselves if this is impractical) are
available open source from the \swift{} repository.

\subsection{Sod shock tube}
\label{sec:sodshock}

\begin{figure}
    \centering
    \includegraphics{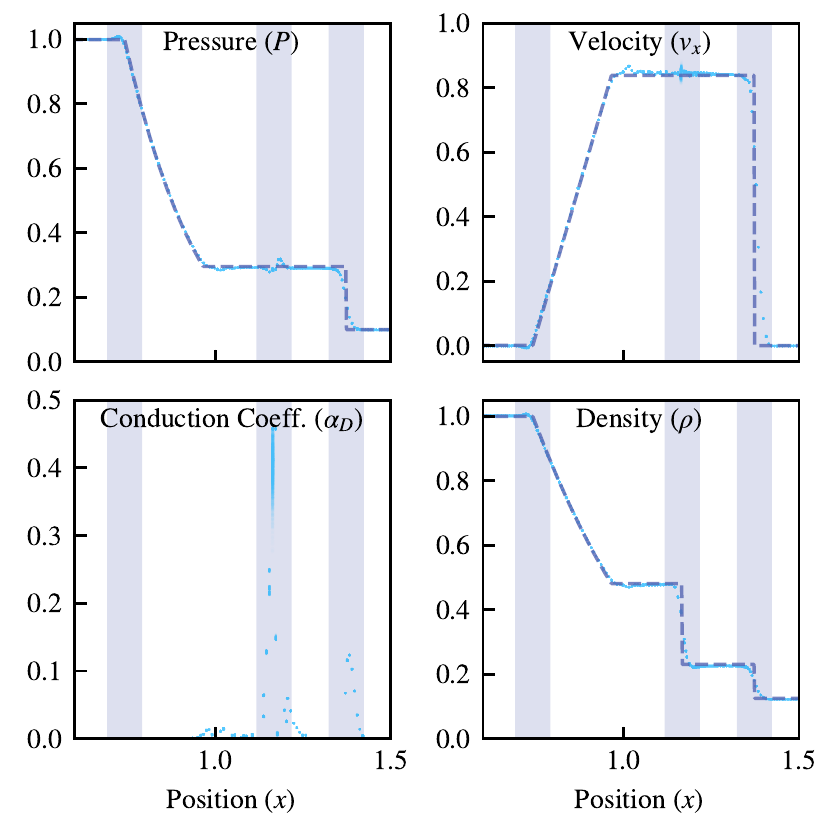}
    \caption{Individual quantities plotted against the analytic solution
    (purple dashed line) for the Sod shock tube in 3D. The horizontal
    axis shows the $x$ position of the particles. All particles are shown in
    blue, with the purple shading in the background showing the regions
    considered for the convergence (Fig. \ref{fig:sodconverge}) with the
    rarefaction wave, contact discontinuity, and shock, shown from left to
    right. All panels are shown at the same time $t=0.2$, and for the same
    resolution level, using the $64^3$ and $128^3$ initial conditions for
    $x < 1$ and $x > 1$ respectively.}
    \label{fig:sodshock}
\end{figure}

The \citet{Sod1978} shock tube is a classic Riemann problem often used to
test hydrodynamics codes. The tube is made up of three main sections in the
final solution : the rarefaction wave (between $0.7 < x < 1.0$), contact
discontinuity (at position $x\approx1.2$), and a weak shock (at position
$x\approx1.4$) at the time that we show it in Figure \ref{fig:sodshock}.

\subsubsection{Initial Conditions}

The initial conditions for the Sod shock tube uses body centred cubic
lattices to ensure maximally symmetric lateral forces in the initial state.
Two lattices with equal particle masses, one at a higher density by a factor
of 8 (e.g. one with $32^3$ particles and one with $64^3$ particles) are
attached at $x=1.0$\footnote{This simplistic particle arrangement does cause
a slight problem at the interface at higher (i.e. greater than one)
dimensions. In 3D, some particles may have spurious velocities in the $y$ and
$z$ directions at the interface, due to asymmetries in the neighbours found
on the left and right side of the boundary. To offset this, the lattices are
placed so that the particles are aligned along the x-axis wherever possible
over the interface, however some spurious forces still result. }. This forms
a discontinuity, with the higher density lattice being placed on the left
with $\rho_L = 1$ and the lower density lattice on the right with $\rho_R =
1/8$. The velocities are initially set to zero for all particles and
pressures set to be $P_L = 1$ and $P_R = 0.1$.

\subsubsection{Results}

\begin{figure}
    \centering
    \includegraphics{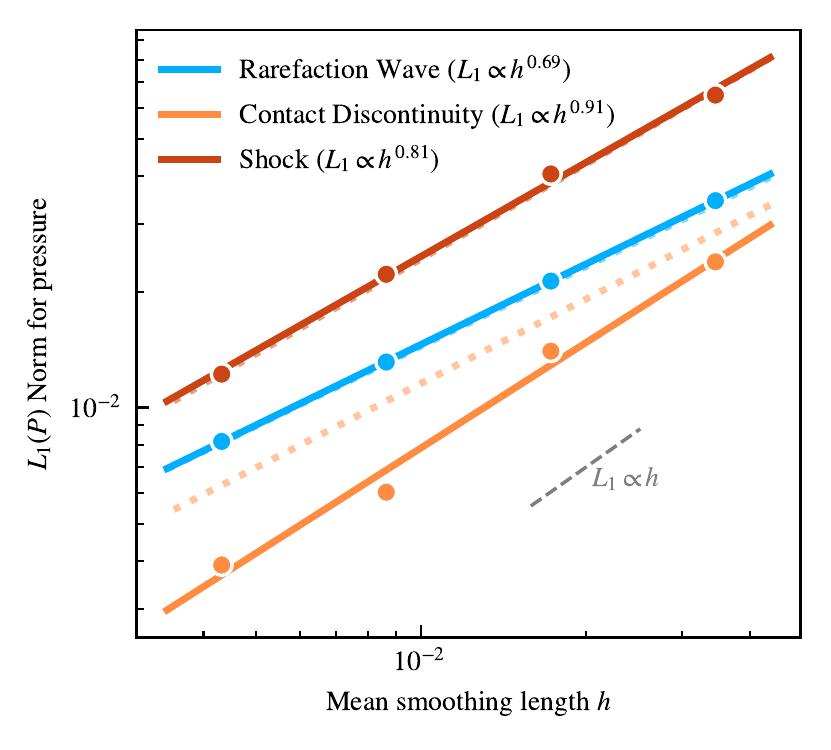}
    \vspace{-0.2cm}
    \caption{Pressure convergence for the three regions in Fig
    \ref{fig:sodshock}. The solid lines show fits to the data at various
    resolution levels (points) for each region, with the dotted lines showing
    convergence speed when the artificial conduction term is removed. The
    dashed grey line shows the expected speed of convergence for shocks in
    SPH simulations, to guide the eye, with a dependence of $L_1 \propto h$.}
    \label{fig:sodconverge}
\end{figure}

Fig. \ref{fig:sodshock} shows the shock tube at $t=1$, plotted against the
analytic solution. This figure shows the result from the $64^3$ and $128^3$
initial condition. In general the simulation data (blue points) shows very
close agreement with the analytic solution (purple dashed line).

The three purple bands correspond to three distinct regions within the shock
tube. The furthest left is the rarefaction wave, which is an adiabatically
expanding fluid. The band covers the turnover point of the wave, as this is where
the largest deviation from the analytic solution is present. There is a slight
overestimation of the density at this turnover point, primarily due to the 
symmetric nature of the SPH kernel.

The next band shows the contact discontinuity. No effort is made to suppress
this discontinuity in the initial conditions (i.e. they are not relaxed).
There is a small pressure blip, of a similar size to that seen with schemes
employing Riemann solvers such as GIZMO \citep{Hopkins2015}. There is no
large velocity discontinuity associated with this pressure blip as is seen
with SPH schemes that do not explicitly treat the contact discontinuity (note
that every particle present in the simulation is shown here) with some form
of conduction, a smoothed pressure field, or other method. Due to the strong
discontinuity in internal energy present in this region, the artificial
conduction coefficient $\alpha_D$ peaks, allowing for the pressure `blip' to
be reduced to one with a linear gradient.

The final section of the system, the rightmost region, is the shock. This
shock is well captured by the scheme. There is a small activation of the
conduction coefficient in this region, which is beneficial as it aids in
stabilising the shock front \citep{Hu2014}. This shows that the conduction
limiter (\S \ref{sec:conductionlimiter}) does not eliminate this beneficial
property of artificial conduction within these frequently present weak
(leading to $\alpha_V \lesssim 1.0$) shocks.

In an ideal case, the scheme would be able to converge
at second order $L_1 \propto h^2$ away from shocks, and at first order $L_1
\propto h$ within shocks \citep{Price2018}. Here the $L_1$ norm of a band
is defined as
\begin{equation}
    L_1(K) = \frac{1}{n}\sum_n |K_{\rm sim}(\vec{x}) - K_{\rm ref}(\vec{x})|
    \label{eqn:L1}
\end{equation}
with $K$ some property of the system such as pressure, the subscripts
sim and ref referring to the simulation data and reference solution respectively,
and $n$ the number of particles in the system.

Fig. \ref{fig:sodconverge} shows the convergence properties of the \sphenix{}
scheme on this problem, using the pressure field in this case as the convergence
variable. Compared to a scheme without artificial conduction (dotted lines), the
\sphenix{} scheme shows significantly improved convergence and a lower norm in
the contact discontinuity, without sacrificing accuracy in other regions.
\subsection{Sedov-Taylor Blastwave}

\begin{figure}
    \centering
    \includegraphics{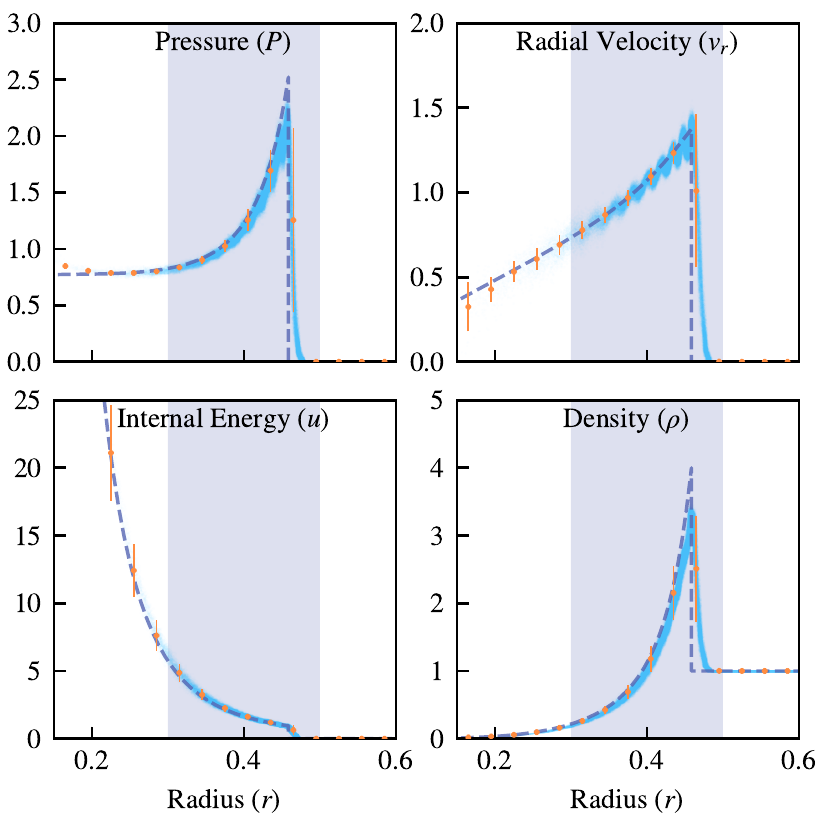}
    \caption{Particle properties at $t=0.1$ shown against the analytic
    solution (purple dashed line) for the Sedov-Taylor blastwave. A random
    sub-set of 1/5th of the particles are shown in blue, with the orange
    points showing the mean value within equally spaced horizontal bins with
    one standard deviation of scatter. The background purple band shows the
    region considered for measuring convergence in Fig.
    \ref{fig:sedovconvergence}. This figure shows the results for a $128^3$
    particle glass file.}
    \label{fig:sedov}
\end{figure}

The Sedov-Taylor blastwave \citep[Sedov blast;][]{Taylor1950, Sedov1959}
follows the evolution of a strong shock front through an initially isotropic
medium. This is a highly relevant test for cosmological simulations, as this
is similar to the implementations used for sub-grid (below the resolution
scale) feedback from stars and black holes. In SPH schemes this
effectively tests the artificial viscosity scheme for energy conservation; if
the scheme does not conserve energy the shock front will be misplaced.

\subsubsection{Initial Conditions}

Here, we use a glass file generated by allowing a uniform grid of particles
to settle to a state where the kinetic energy has stabilised. The particle
properties are then initially set such that they represent a gas with
adiabatic index $\gamma = 5/3$, a uniform pressure of $P_0 = 10^{-6}$,
density $\rho_0 = 1$, all in a 3D box of side-length 1. Then, the $n=15$
particles closest to the centre of the box have energy $E_0 = 1/n$ injected
into them. This corresponds, roughly, to a temperature jump of a factor of
$\sim 10^5$ over the background medium.

\subsubsection{Results}

\begin{figure}
    \centering
    \includegraphics{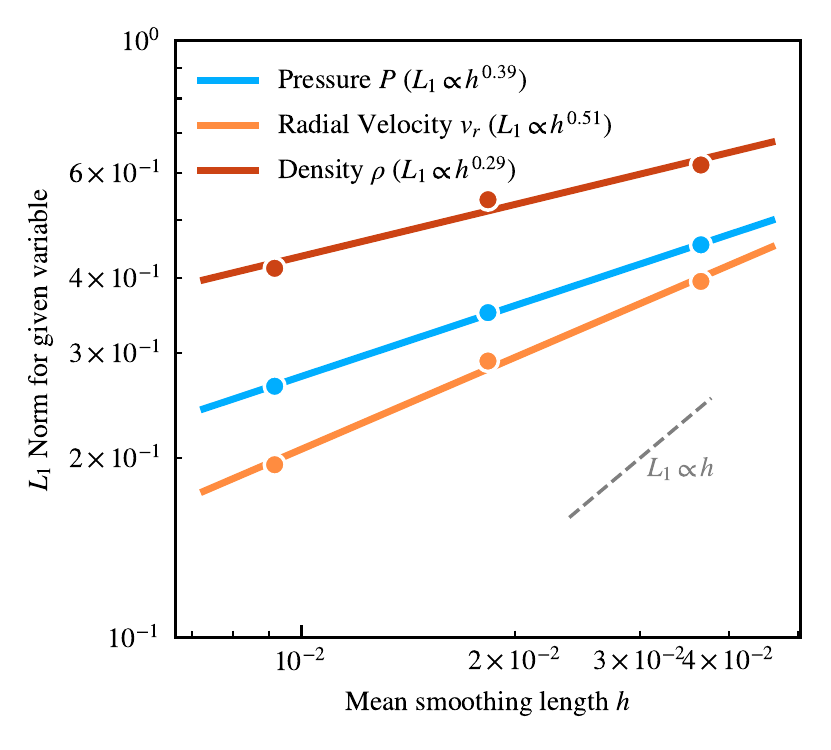}
    \vspace{-0.2cm}
    \caption{$L_1$ Convergence with mean smoothing length for various
    particle fields in the Sedov-Taylor blastwave test, measured at $t=0.1$
    against the analytic solution within the purple band of Fig.
    \ref{fig:sedov}. Each set of points shows a measured value from an
    individual simulation, with the lines showing a linear fit to the data in
    logarithmic space. Dotted lines for the simulation without conduction are
    not shown as they lie exactly on top of the lines shown here.}
    \label{fig:sedovconvergence}
\end{figure}

Fig. \ref{fig:sedov} shows the particle properties of the highest resolution
initial condition ($128^3$) at $t=0.1$ against the analytic solution. The
\sphenix{} scheme closely matches the analytic solution in all particle
fields, with the only deviation (aside from the smoothed shock front, an
unavoidable consequence of using an SPH scheme) being a slight upturn in
pressure in the central region (due to a small amount of conduction in this
region). Of particular note is the position of the shock front matching
exactly with the analytic solution, showing that the scheme conserves energy
in this highly challenging situation thanks to the explicitly symmetric
artificial viscosity equation of motion. The \sphenix{} scheme shows
qualitatively similar results to the \phantomsph{} scheme on this problem
\citep{Price2018}.

SPH schemes in general struggle to show good convergence on shock problems
due to their inherent discontinuous nature. Ideal convergence for shocks
with the artificial viscosity set-up used in \sphenix{} is only first order
(i.e. $L_1 \propto h$).

Fig. \ref{fig:sedovconvergence} shows the $L_1$ convergence for various
fields in the Sedov-Taylor blastwave as a function of mean smoothing length.
Convergence here has a best-case of $L_1(v) \propto h^{1/2}$ in real terms,
much slower than the expected $L_1 \propto h^{-1}$. This is primarily due to
the way that the convergence is measured; the shock front is not resolved
instantaneously (i.e. there is a rise in density and velocity over some small
distance, reaching the maximum value at the true position) at the same
position as in the analytic solution. However, all resolution levels show an
accurately placed shock front and a shock width that scales linearly with
resolution (see Appendix \ref{app:sedovblast} for more information).
\subsection{Gresho-Chan Vortex}
\label{sec:gresho}

\begin{figure}
    \centering
    \includegraphics{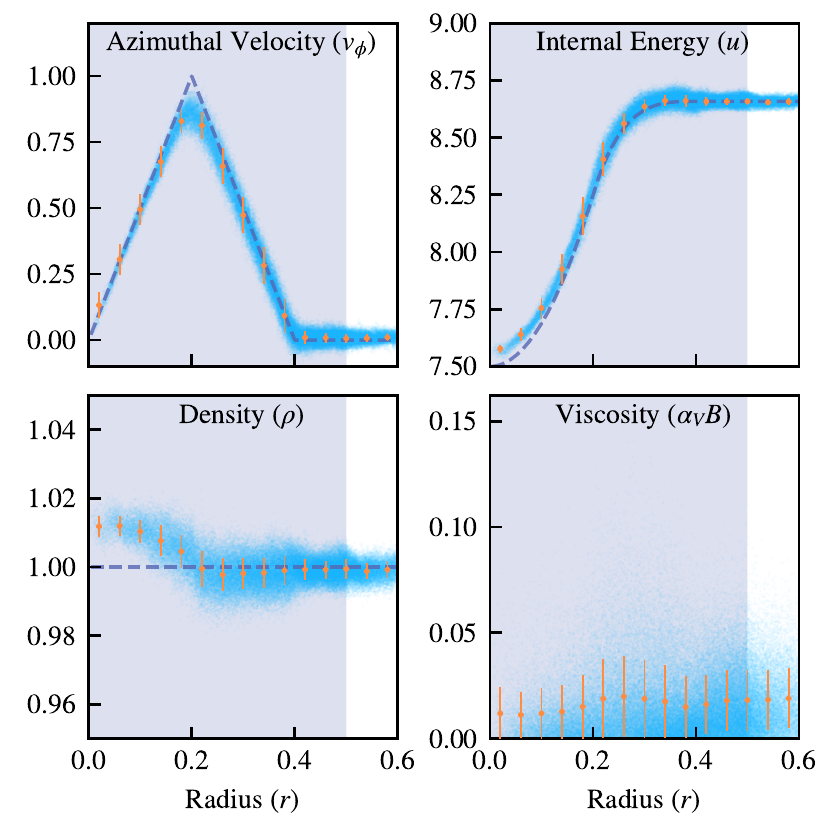}
    \caption{Gresho vortex at $t=1.3$ after one rotation of the vortex peak
    with the \sphenix{} scheme using a background resolution of $512^2$ and
    with a mach number of $\mathcal{M} = 0.33$. Here the blue points show all
    particles in the volume, the purple band the region used for convergence
    testing in Fig. \ref{fig:greshoconvergence}, and the purple dashed line shows
    the analytic solution. The viscosity switch panel shows a very low maximal
    value (0.15) relative to the true maximum allowed by the code ($\alpha_V B =
    2.0$), with the mean value (orange points with error bars indicating
    one standard deviation of scatter) of around 0.02 showing an
    excellent activation of the viscosity reducing switches throughout the
    \sphenix{} scheme.}
    \label{fig:gresho}
\end{figure}

The Gresho-Chan vortex \citep{Gresho1990a} is typically used to test for the
conservation of vorticity and angular momentum, and is performed here in two
dimensions. Generally, it is expected that the performance of SPH on this test
is more dependent on the kernel employed \citep[see][]{Dehnen2012}, as long as a
sufficient viscosity-suppressing switch is used.

\subsubsection{Initial Conditions}

The initial conditions use a two dimensional glass file, and treat the gas
with an adiabatic index $\gamma = 5/3$, constant density $\rho_0 = 1$, in a
square of side-length 1. The particles are given azimuthal velocity
\begin{equation}
    v_\phi = \begin{cases}
        5r & r < 0.2 \\
        2 - 5r & 0.2 \leq r < 0.4 \\
        0 & r \geq 0.4 
    \end{cases}
\end{equation}
with the pressure set so that the system is in equilibrium as
\begin{equation}
    P_0 = \begin{cases}
        5 + 12.5 r^2 & r < 0.2 \\
        9 + 12.5 r^2 - 20 r + 4 \log(5 r) & 0.2 \leq r < 0.4 \\
        3 + 4\log(2) & r \geq 0.4 
    \end{cases}
\end{equation}
where $r = \sqrt{x^2 + y^2}$ is the distance from the box centre.

\subsubsection{Results}

Fig. \ref{fig:gresho} shows the state of a high resolution (using a glass
containing $512^2$ particles) result after one full rotation at the peak of
the vortex ($r=0.2$, $t=1.3$). The vortex is well supported, albeit with some
scatter, and the peak of the vortex is preserved. There has been some
transfer of energy to the centre with a higher density and internal energy
than the analytic solution due to the viscosity switch (shown on the bottom
right) having a very small, but nonzero, value. This then allows for some of
the kinetic energy to be transformed to thermal, which is slowly transported
towards the centre as this is the region with the lowest thermal pressure.

\begin{figure}
    \centering
    \includegraphics{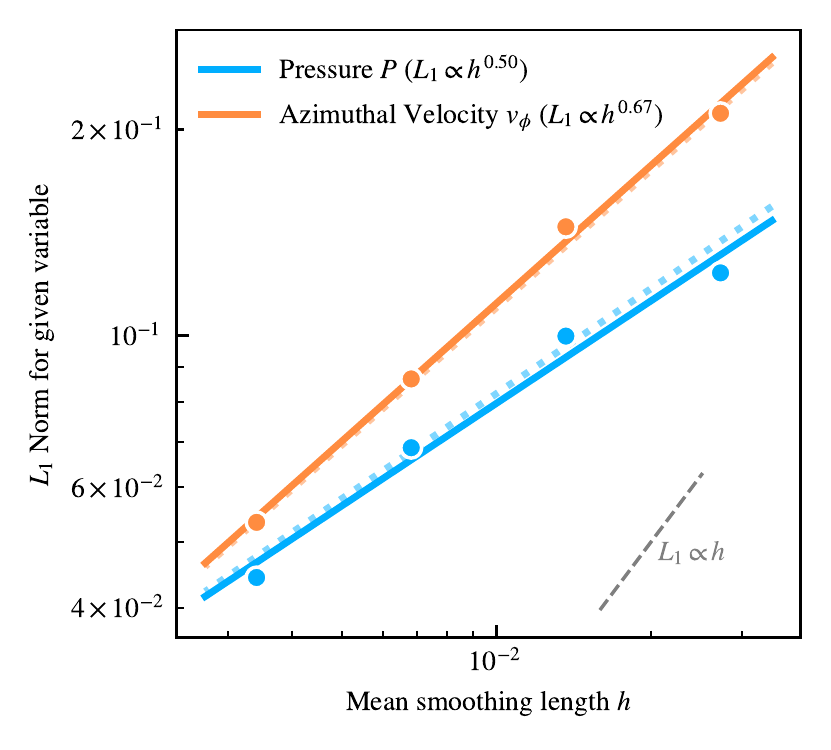}
    \vspace{-0.2cm}
    \caption{$L_1$ Convergence with mean smoothing length for various
    particle fields in the Gresho vortex test, measured against the analytic
    solution within the shaded region of Fig. \ref{fig:gresho}. Each set of
    points shows a measured value from an individual simulation, with the
    lines showing a linear fit to the data in logarithmic space. The solid
    lines show results obtained with the full \sphenix{} scheme, with dotted
    lines showing the results with the artificial conduction scheme
    disabled.}
    \label{fig:greshoconvergence}
\end{figure}

Fig. \ref{fig:greshoconvergence} shows the convergence properties for the
vortex, with the \sphenix{} scheme providing convergence as good as
$L_1\propto h^{0.7}$ for the azimuthal velocity. As there are no non-linear
gradients in internal energy present in the simulation there is very little
difference between the simulations performed with and without conduction at
each resolution level due to the non-activation of Eqn. \ref{eqn:alphaDevo}.
This level of convergence is similar to the the rate seen in \citet{Dehnen2012}
implying that the \sphenix{} scheme, even with its less complex viscosity
limiter, manages to recover some of the benefits of the more complex
\emph{Inviscid} scheme thanks to the novel combination of switches employed.
\subsection{Noh Problem}
\label{sec:noh}

\begin{figure}
    \centering
    \includegraphics{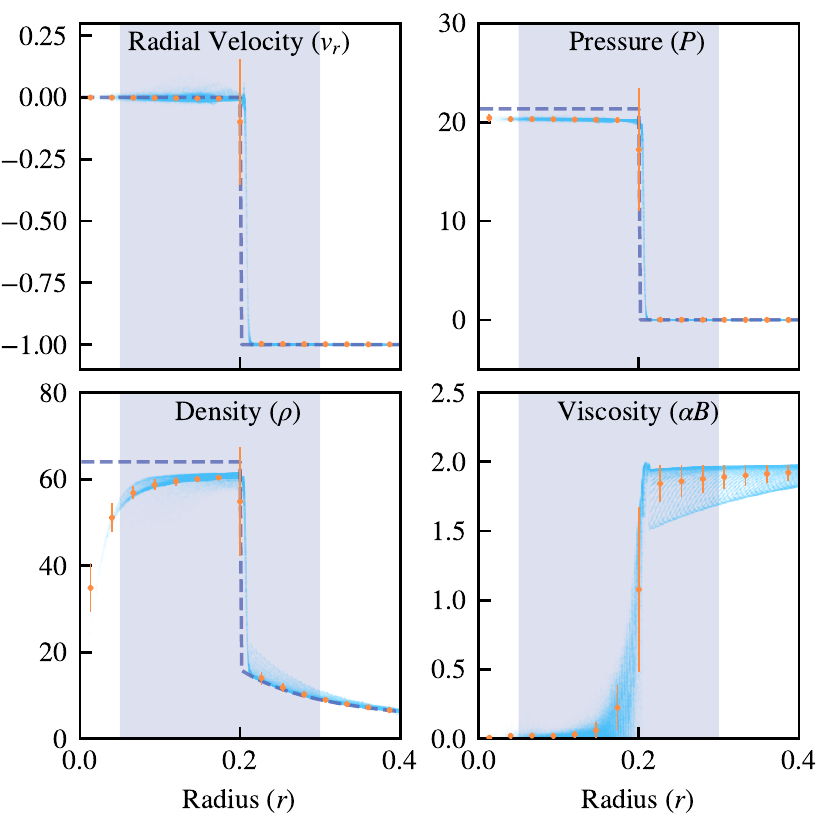}
    \caption{Noh problem simulation state at $t=0.6$, showing a random
    sub-set of 1/100th of all of the particles plotted as blue points, the
    analytical solution as a dashed purple line, and binned quantities as
    orange points with error bars showing one standard deviation of scatter
    in that bin. The background shaded band shows the region considered for
    convergence in Fig. \ref{fig:nohconv}, with this figure showing the
    highest resolution simulation performed, using $512^3$ particles. This
    simulation state is also visualised in Fig.
    \ref{fig:nohimg}.}
    \label{fig:noh}
\end{figure}

The \citet{Noh1987} problem is known to be extremely challenging,
particularly for particle-based codes, and generally requires a high particle
number to correctly capture due to an unresolved convergence point. It tests
a converging flow that results in a strong radial shock. This is an extreme,
idealised, version of an accretion shock commonly present within galaxy
formation simulations.

\subsubsection{Initial Conditions}

There are many ways to generate initial conditions, from very simple schemes
to schemes that attempt to highly optimise the particle distribution
\citep[see e.g.][]{Rosswog2020a}. Here, we use a simple initial condition,
employing a body-centred cubic lattice distribution of particles in a
periodic box. The velocity of the particles is then set such that there is a
convergent flow towards the centre of the box,
\begin{equation}
    \vec{v} = - \frac{\vec{C} - \vec{x}}{|\vec{C} - \vec{x}|}
\end{equation}
with $\vec{C} = 0.5 L (1, 1, 1)$, where $L$ is the box side-length,
the coordinate at the centre of the volume. This gives every
particle a speed of unity, meaning those in the centre will have
extremely high relative velocities. We cap the minimal
value of $|\vec{C} - \vec{x}|$ to be $10^{-10} L$ to prevent
a singularity at small radii.

The simulation is performed in a dimensionless co-ordinate system,
with a box-size of $L=1$.

\subsubsection{Results}

\begin{figure}
    \centering
    \includegraphics{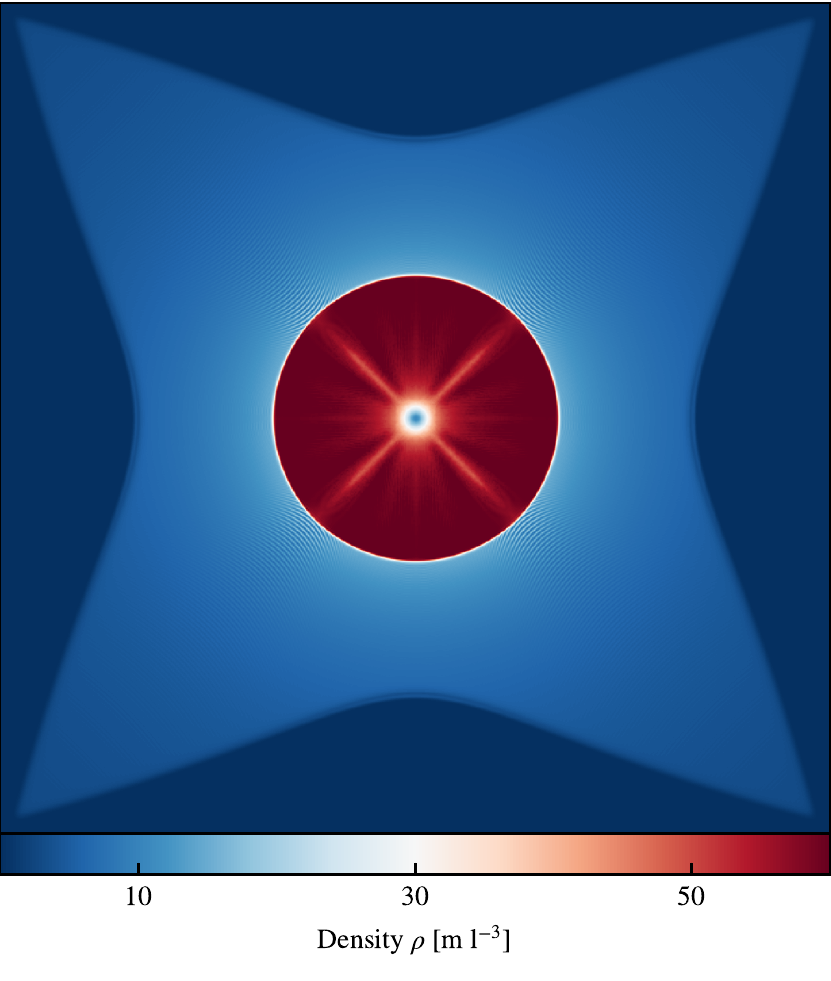}
    \vspace{-0.2cm}
    \caption{A density slice through the centre of the Noh probeem at $t=0.6$
    corresponding to the particle distribution shown in Fig. \ref{fig:noh}.
    The \sphenix{} scheme yields almost perfect spherical symmetry for the
    shock, but does not capture the expected high density in the central
    region, likely due to lower than required artificial conductivity (see
    Appendix \ref{app:nohcond} for more information).}
    \label{fig:nohimg}
\end{figure}

\begin{figure}
    \centering
    \includegraphics{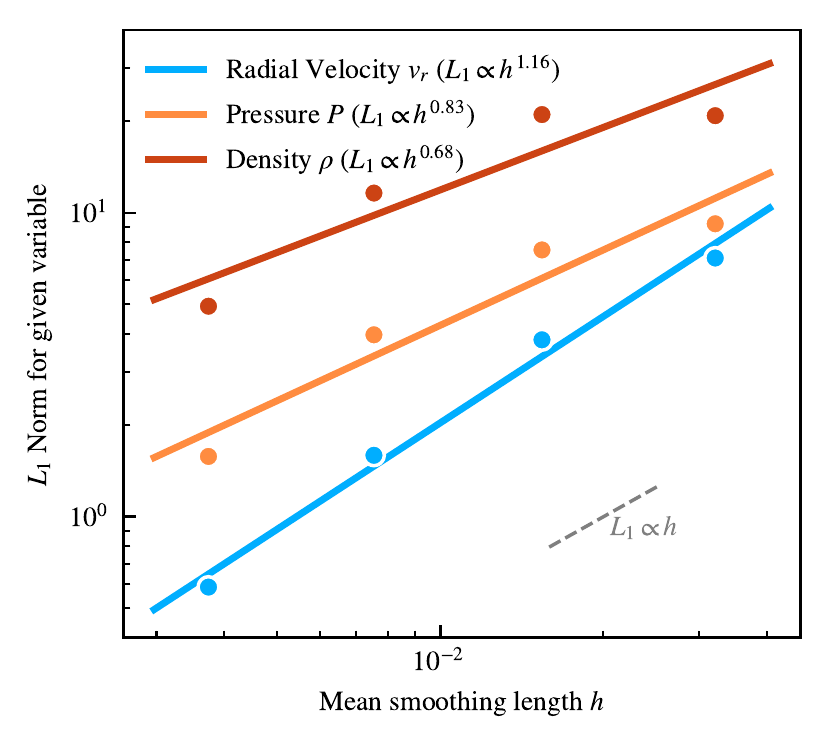}
    \caption{$L_1$ convergence test for various particle properties
    at $t=0.6$ for the Noh problem, corresponding to the particle
    distribution shown in Fig. \ref{fig:noh}. The lines without
    conduction are not shown here as there is little difference
    between the with and without conduction case, due to the
    extremely strong shock present in this test (leading
    to low values of the viscosity alpha,
    Equation \ref{eqn:conductionlimiter}).}
    \label{fig:nohconv}
\end{figure}

The state of the simulation is shown at time $t=0.6$ in Fig. \ref{fig:noh} and
visualised in Fig. \ref{fig:nohimg}, which shows the radial velocity, which
should be zero inside of the shocked region (high density in Fig.
\ref{fig:nohimg}), and the same as the initial conditions (i.e. -1
everywhere) elsewhere. This behaviour is captured well, with a
small amount of scatter, corresponding to the small radial variations
in density shown in the image.

The profile of density as a function of radius is however less well captured,
with some small waves created by oscillations in the artificial viscosity
parameter \citep[see e.g.][for a scheme that corrects for these
errors]{Rosswog2020}. This can also be seen in the density slice, and is a
small effect that also is possibly exacerbated by our non-perfect choice of
initial conditions, but is also present in the implementation shown in
\citet{Rosswog2020a}. The larger, more significant, density error is shown
inside the central part of the shocked, high-density, region. This error is
ever-present in SPH schemes, and is likely due to both a lack of artificial
conduction in this central region \citep[as indicated by][note the excess
pressure in the centre caused by `wall heating']{Noh1987} and the unresolved
point of flow convergence.

The Noh problem converges well using \sphenix{}, with better than linear
convergence for the radial velocity (Fig. \ref{fig:nohconv}; recall that
for shocks SPH is expected to converge with $L_1 \propto h$).

This problem does not activate the artificial conduction in the \sphenix{}
implementation because of the presence of Equation
\ref{eqn:conductionlimiter} reducing conductivity in highly viscous flows, as
well as our somewhat conservative choice for artificial conduction
coefficients (see Appendix \ref{app:nohcond} for more details on this topic).
However, as these are necessary for the practical functioning of the
\sphenix{} scheme in galaxy formation simulations, and due to this test being
highly artificial, this outcome presents little concern.
\subsection{Square Test}
\label{sec:squaretest}

\begin{figure}
    \centering
    \includegraphics{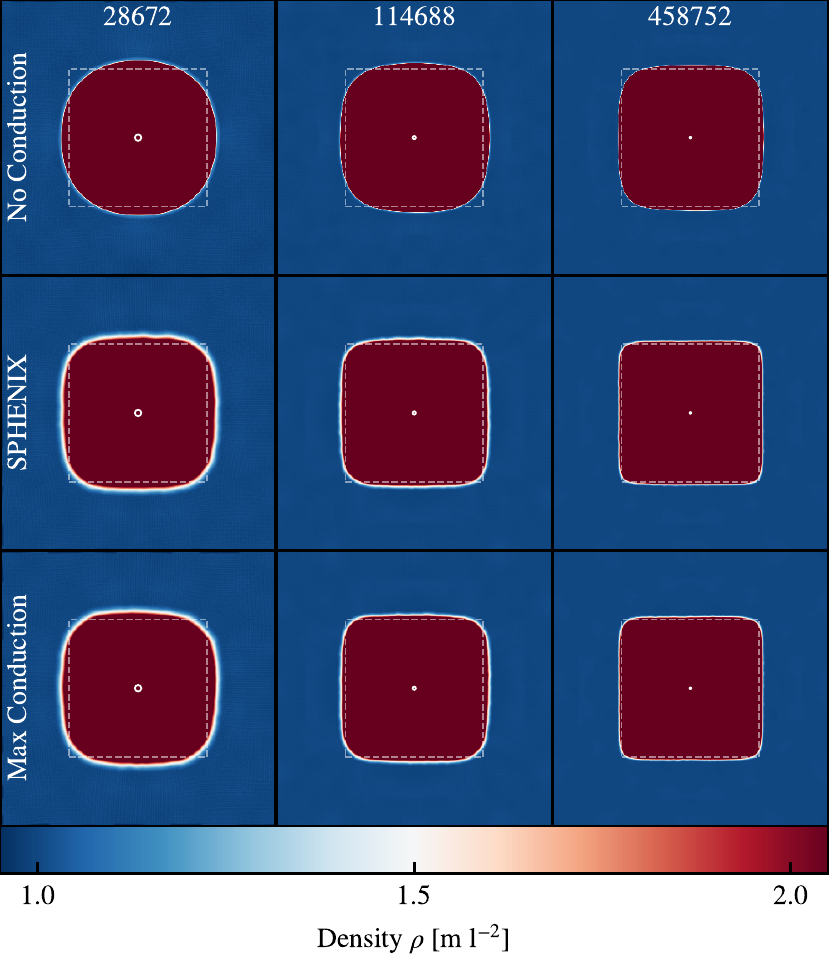}
    \caption{The density field for the square test at $t=4$, shown at various
    resolution levels (different columns, numbers at the top denote the number of
    particles in the system) and with various modifications to the underlying
    SPH scheme (different rows). The dashed line shows the initial boundary of
    the square that would be maintained with a perfect scheme due to the
    uniform pressure throughout. The white circle at the centre of the square
    shows a typical smoothing length for this resolution level. Vertically,
    the scheme with no conduction is shown at the top, with the \sphenix{}
    scheme in the middle and a scheme with the conduction coefficient set to
    the maximum level throughout at the bottom. The schemes with conduction
    maintain the square shape significantly better than the one without
    conduction, and the \sphenix{} limiters manage to provide the appropriate
    amount of conduction to return to the same result as the maximum
    conduction case.}
    \label{fig:squareconvergence}
\end{figure}

The square test, first presented in \citet{Saitoh2013}, is a particularly
challenging test for schemes like \sphenix{} that do not use a smoothed
pressure in their equation of motion, as they typically lead to an artificial
surface tension at contact discontinuities (the same ones that lead to the
pressure blip in \S \ref{sec:sodshock}). This test is a more challenging
variant of the ellipsoid test presented in \citet{Hess2010}, as the square
includes sharp corners which are more challenging for conduction schemes to
capture.

\subsubsection{Initial conditions}

The initial conditions are generated using equal mass particles. We set up a
grid in 2D space with $n\times n$ particles, in a box of size $L=1$. The
central $0.5 \times 0.5$ square is set to have a density of $\rho_C = 4.0$,
and so is replaced with a grid with $2n \times 2n$ particles, with the outer
region having $\rho_O = 1.0$. The pressures are set to be equal with $P_C =
P_O = 1.0$, with this enforced by setting the internal energies of the
particles to their appropriate values. All particles are set to be completely
stationary in the initial conditions with $\vec{v} = 0$. The initial conditions
are not allowed to relax in any way.

\subsubsection{Results}

Fig. \ref{fig:squareconvergence} shows the square test at $t=4$ for four
different resolution levels and three different variations on the \sphenix{}
scheme. By this time the solutions are generally very stable. The top row
shows the \sphenix{} scheme without any artificial conduction enabled (this
is achieved by setting $\alpha_{D, {\rm max}}$ to zero) and highlights the
typical end state for a Density-Energy SPH scheme on this problem. Artificial
surface tension leads to the square deforming and rounding to become more
circular.

The bottom row shows the \sphenix{} scheme with the artificial conduction
switch removed; here $\alpha_{D, {\rm min}}$ is set to the same value as
$\alpha_{D, {\rm max}} = 1$. The artificial conduction scheme significantly
reduces the rounding of the edges, with a rapid improvement as resolution
increases. The rounding present here only occurs in the first few steps as
the energy outside the square is transferred to the boundary region to
produce a stable linear gradient in internal energy.

Finally, the central row shows the \sphenix{} scheme, which gives a result
indistinguishable from the maximum conduction scenario. This is despite the
initial value for the conduction coefficient $\alpha_D = 0$, meaning it must
ramp up rapidly to achieve such a similar result. The \sphenix{} result here
shows that the choices for the conduction coefficients determined from the
Sod tube (\S \ref{sec:sodshock}) are not only appropriate for that test, but
apply more generally to problems that aim to capture contact
discontinuities.

\subsection{2D Kelvin-Helmholtz Instability}
\label{sec:kelvinhelmholtz}

\begin{figure}
    \centering
    \includegraphics{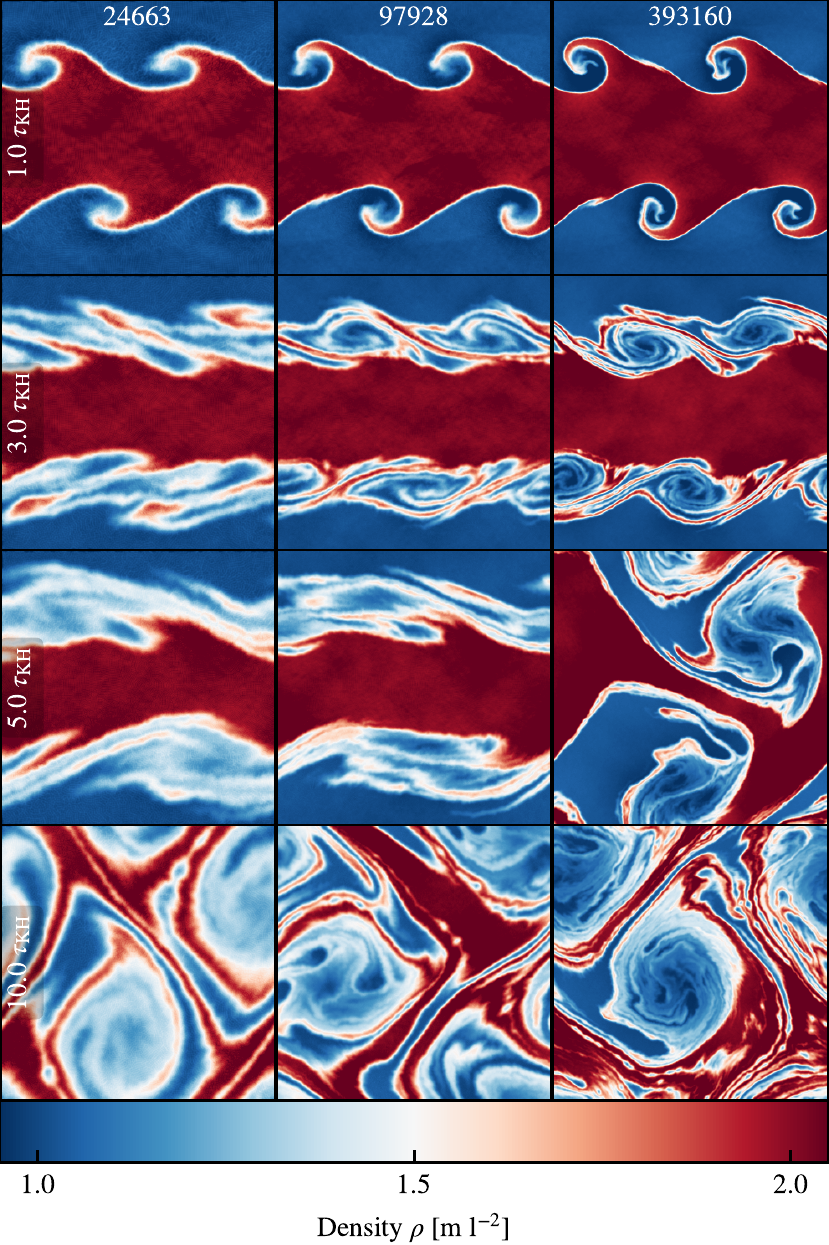}
    \vspace{-0.2cm}
    \caption{Density map of the standard Kelvin-Helmholtz 2D test at
    various resolutions (different columns, with the number of particles in
    the volume at the top) and at various times (different rows showing times
    from $t=\tau_{\rm KH}$ to $t=10\tau_{\rm KH}$). Despite this being
    a challenging test for SPH, the instability is captured well at all 
    resolutions, with higher resolution levels capturing finer
    details.}
    \label{fig:kelvinhelmholtz}
\end{figure}

The two dimensional Kelvin-Helmholtz instability is presented below. This
test is a notable variant on the usual Kelvin-Helmholtz test as it includes a
density jump at constant pressure (i.e. yet another contact discontinuity).
This version of the Kelvin-Helmholtz instability is performed in two
dimensions. A recent, significantly more detailed, study of Kelvin-Helmholtz
instabilities within SPH is available in \citet{Tricco2019a}. In this section 
we focus on qualitative comparisons and how the behaviour of the instability
changes with resolution within \sphenix{}.

\subsubsection{Initial conditions}

The initial conditions presented here are similar to those in
\citet{Price2008}, where they discuss the impacts more generally of the
inclusion of artificial conduction on fluid mixing instabilities. This is set
up in a periodic box of length $L=1$, with the central band between $0.25 < y
< 0.75$ set to $\rho_C = 2$ and $v_{C, x} = 0.5$, with the outer region
having $\rho_O = 1$ and $v_{O, x} = -0.5$ to set up a shear flow. The
pressure $P_C = P_O = 2.5$ is enforced by setting the internal energies of
the equal mass particles. Particles are initially placed on a grid with equal
separations. This is the most challenging version of this test for SPH
schemes to capture as it includes a perfectly sharp contact discontinuity;
see \citet{Agertz2007} for more information.

We then excite a specific mode of the instability, as in typical SPH simulations
un-seeded instabilities are dominated by noise and are both unpredictable and
unphysical, preventing comparison between schemes.

\subsubsection{Results}

Fig. \ref{fig:kelvinhelmholtz} shows the simulation after various multiples of the
Kelvin-Helmholtz timescale for the excited instability,
with $\tau_{\rm KH}$ given by
\begin{equation}
    \tau_{\rm KH} = \frac{(1 + \chi)\lambda}{\bar{v} \sqrt{\chi}}
    \label{eqn:taukh}
\end{equation}
where $\chi = \rho_C / \rho_O = 2$ is the density contrast, $\bar{v} = v_{I,
x} - v_{O, x} = 1$ the shear velocity, and $\lambda = 0.5$ the wavelength of
the seed perturbation along the horizontal axis \citep[e.g][]{Hu2014}. The
figure shows three initial resolution levels, increasing from left to right.
Despite this being the most challenging version of the Kelvin-Helmholtz test
(at this density contrast) for a Density-Energy based SPH scheme, the
instability is captured well at all resolutions, with higher resolutions
allowing for more rolls of the `swirl' to be captured. In particular, the
late-time highly mixed state shows that with the conduction removed after a
linear gradient in internal energy has been established, the \sphenix{}
scheme manages to preserve the initial contact discontinuity well. Due to
the presence of explicit artificial conduction, \sphenix{} seems to diffuse more
than other schemes on this test \citep[e.g.][]{Hu2014,Wadsley2017}, leading
to the erausre of some small-scale secondary instabilities.

The non-linear growth rate of the swirls is resolution dependent within this
test, with higher-resolution simulations showing faster growth of the
largest-scale modes. This is due to better capturing of the energy initially
injected to perturb the volume to produce the main instability, with higher
resolutions showing lower viscous losses. 

\begin{figure}
    \centering
    \includegraphics{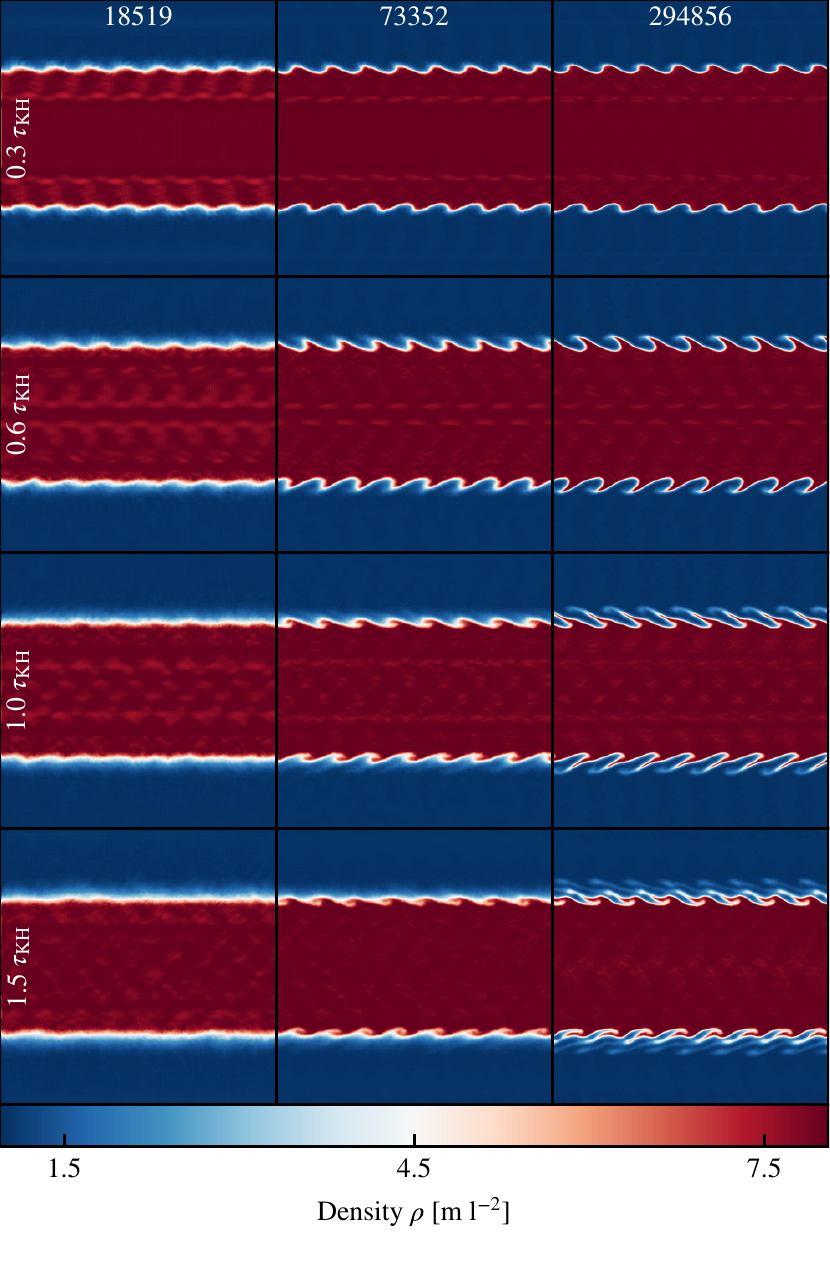}
    \vspace{-0.2cm}
    \caption{The same as Fig. \ref{fig:kelvinhelmholtz}, but this time using
    initial conditions with a significantly higher (1:8 instead of 1:2) density
    contrast. The initial instabilities are captured well at all resolution levels,
    but at the lowest level they are rapidly mixed by the artificial conduction scheme
    due to the lack of resolution elements in the low-density region.}
    \label{fig:kelvinhelmholtzhighdens}
\end{figure}

Fig. \ref{fig:kelvinhelmholtzhighdens} shows a different initial condition
where the density contrast $\chi = 8$, four times higher than the one
initially presented. Because SPH is fundamentally a finite mass method, and
we use equal-mass particles throughout, this is a particularly challenging
test as the low-density region is resolved by so few particles. Here we also
excite an instability with a wavelength $\lambda=0.125$, four times smaller
than the one used for the $\chi=2$ test. This value is chosen for two
reasons; it is customary to lower the wavelength of the seeded instability as
the density contrast is increased when grid codes perform this test as it
allows each instability to be captured with the same number of cells at a
given resolution level; and also to ensure that this test is as challenging
as is practical for the scheme.

\sphenix{} struggles to capture the instability at very low resolutions
primarily due to the lack of particles in the low-density flow \citep[an
issue also encountered by][]{Price2008}. In the boundary region the
artificial conduction erases the small-scale instabilities on a timescale
shorter than their formation timescale (as the boundary region is so large)
and as such they cannot grow efficiently. As the resolution increases,
however, \sphenix{} is able to better capture the linear evolution
of the instability, even capturing turn-overs and the beginning of nonlinear
evolution for the highest resolution.
\subsection{Blob Test}

\begin{figure}
    \centering
    \includegraphics{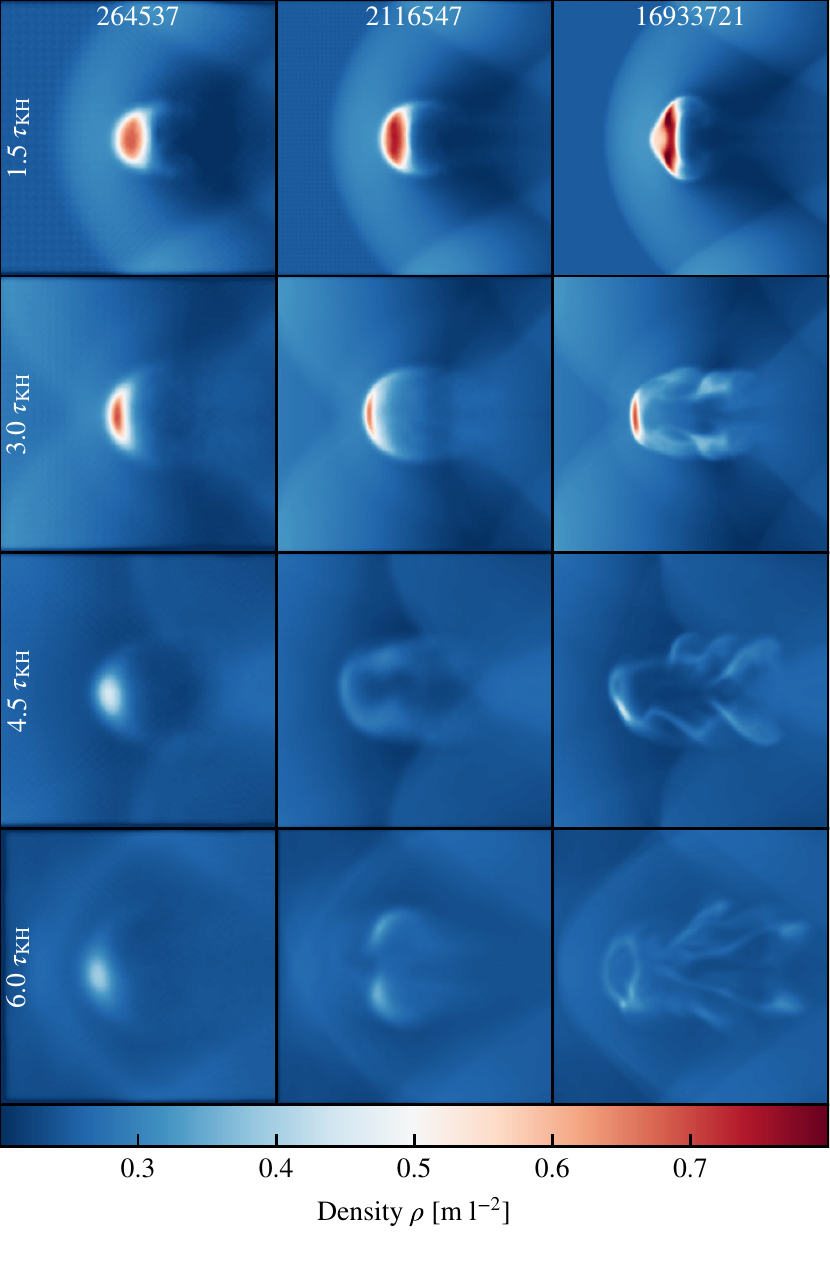}
    \vspace{-0.2cm}
    \caption{Time-evolution of the blob within the supersonic wind at various
    resolution levels (different columns; the number of particles in the
    whole volume is noted at the top) and at various times (expressed as a
    function of the Kelvin-Helmholtz time for the whole blob $\tau_{\rm KH}$;
    different rows). The projected density is shown here to enable all layers
    of the three dimensional structure to be seen. At all resolution levels
    the blob mixes with the surrounding medium (and importantly mixes phases
    with the surrounding medium), with higher resolution simulations
    displaying more thermal instabilities that promote the breaking up of the
    blob.}
    \label{fig:blob}
\end{figure}

The Blob test is a challenging test for SPH schemes \citep[see ][]{Klein1994,
Springel2005} and aims to replicate a scenario where a cold blob of gas falls
through the hot IGM/CGM surrounding a galaxy. In this test, a dense sphere of
cold gas is placed in a hot, low density, and supersonic wind. Ideally, the
blob should break up and dissolve into the wind, but \citet{Agertz2007}
showed that the inability of traditional SPH schemes to exchange entropy
between particles prevents this from occurring. The correct, specific, rate
at which the blob should mix with its surroundings, as well as the structure
of the blob whilst it is breaking up, are unknown.

\subsubsection{Initial Conditions}

There are many methods to set up the initial conditions for the Blob test,
including some that excite an instability to ensure that the blob breaks up
reliably \citep[such as those used in][]{Hu2014}. Here we excite no such
instabilities and simply allow the simulation to proceed from a very basic
particle set-up with a perfectly sharp contact discontinuity. The initial
conditions are dimensionless in nature, as the problem is only specified in
terms of the Mach number of the background medium and the blob density
contrast.

To set up the initial particle distribution, we use two body centred cubic
lattices, one packed at a high-density (for the blob, $\rho_{\rm blob} = 10$)
and one at low density (for the background medium, $\rho_{\rm bg} = 1$). The
low-density lattice is tiled four times in the $x$ direction to make a box of
size $4 \times 1 \times 1$, and at $[0.5, 0.5, 0.5]$ a sphere of radius $0.1$
is removed and filled in with particles from the high-density lattice. The
particles in the background region are given a velocity of $v_{\rm bg} = 2.7$
(with the blob being stationary), and the internal energy of the gas
everywhere is scaled such that the background medium has a mach number of
$\mathcal{M} = 2.7$ and the system is in pressure equilibrium everywhere. 

\subsubsection{Results}

The blob is shown at a number of resolution levels at various times in Fig.
\ref{fig:blob}. At all resolution levels the blob mixes well with the
background medium after a few Kelvin-Helmholtz timescales (see Eqn.
\ref{eqn:taukh} for how this is calculated; here we assume that the
wavelength of the perturbation is the radius of the blob)\footnote{ Note that
here the Kelvin-Helmholtz timescale is 1.1 times the cloud crushing timescale
\citep{Agertz2007}. }. The rate of mixing is consistent amongst all
resolution levels, implying that the artificial conduction scheme is
accurately capturing unresolved mixing at lower resolutions.

The rate of mixing of the blob is broadly consistent with that of modern SPH schemes
and grid codes, however our set of initial conditions appear to mix slightly
slower (taking around $\sim 4-6\tau_{\rm KH}$ to fully mix) than those used
by other contemporary works \citep{Agertz2007, Read2012, Hu2014}, possibly due
to the lack of perturbation seeding \citep[see][Appendix B for more
details]{Read2010a}. When testing these initial conditions
with a scheme that involves a Riemann solver or a Pressure-based scheme (see
Appendix \ref{app:blob}) the rate of mixing is qualitatively similar to the one
presented here. \sphenix{} is unable to fully capture the crushing of the blob
from the centre outwards seen in grid codes and other SPH formulations using
different force expressions \citep{Wadsley2017}, rather preferring to retain a
`plate' of dense gas at the initial point of the blob that takes longer to break
up. A potential explanation for this difference is some residual surface tension
in \sphenix{}. In these highly dynamic situations, it may not be possible for the
artificial conduction to establish a smooth internal energy profile rapidly enough
for small-scale instabilities on the surface to assist in the breakup of the blob.

At low resolutions it is extremely challenging for the method to capture the
break-up of the blob as there are very few particles in the background
medium to interact with the blob due to the factor of 10 density contrast.

\begin{figure*}
    \centering
    \includegraphics{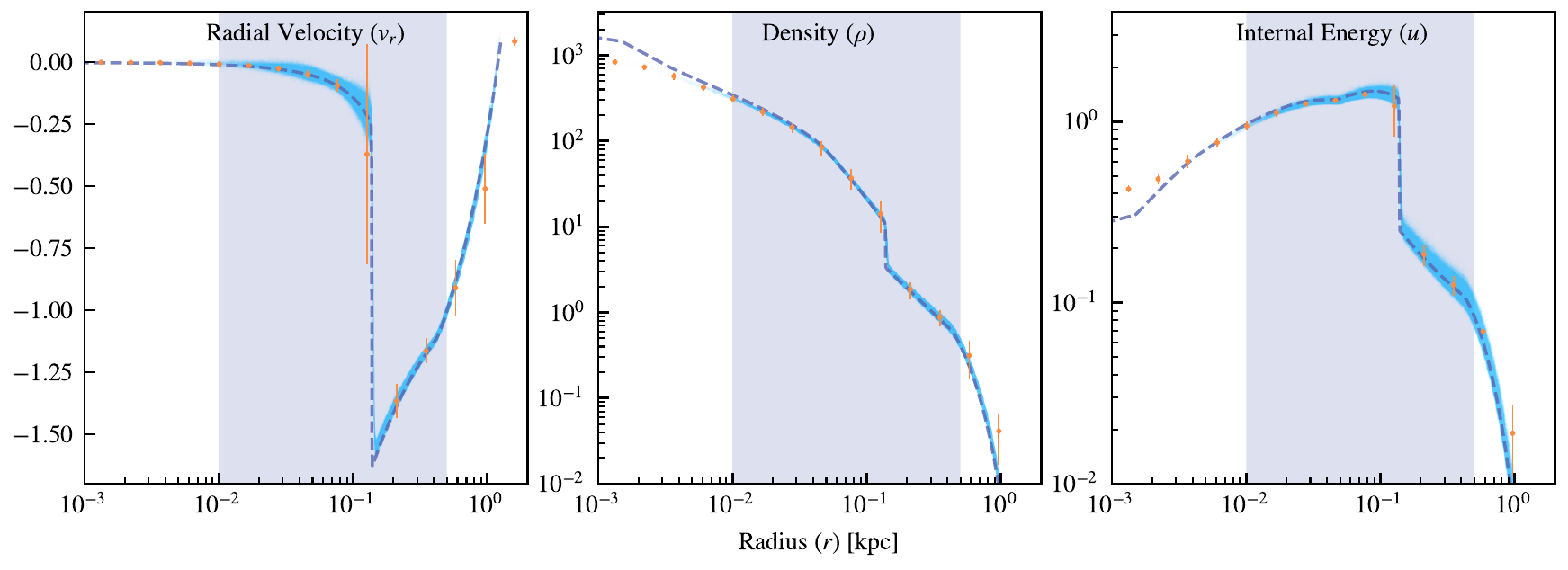}
    \caption{State of the Evrard sphere at $t=0.8$ for a resolution of $10^7$
    particles. A random sub-set of 1/10th of the particles is shown in blue,
    with the solution from a high resolution 1D grid code shown as a purple
    dashed line. The orange points with error bars show the median within a
    radial equally log-spaced bin with the bar showing one standard deviation
    of scatter. The shaded band in the background shows the region considered
    for the convergence test in Fig. \ref{fig:evrardconvergence}.}
    \label{fig:evrard}
\end{figure*}

\begin{figure}
    \centering
    \includegraphics{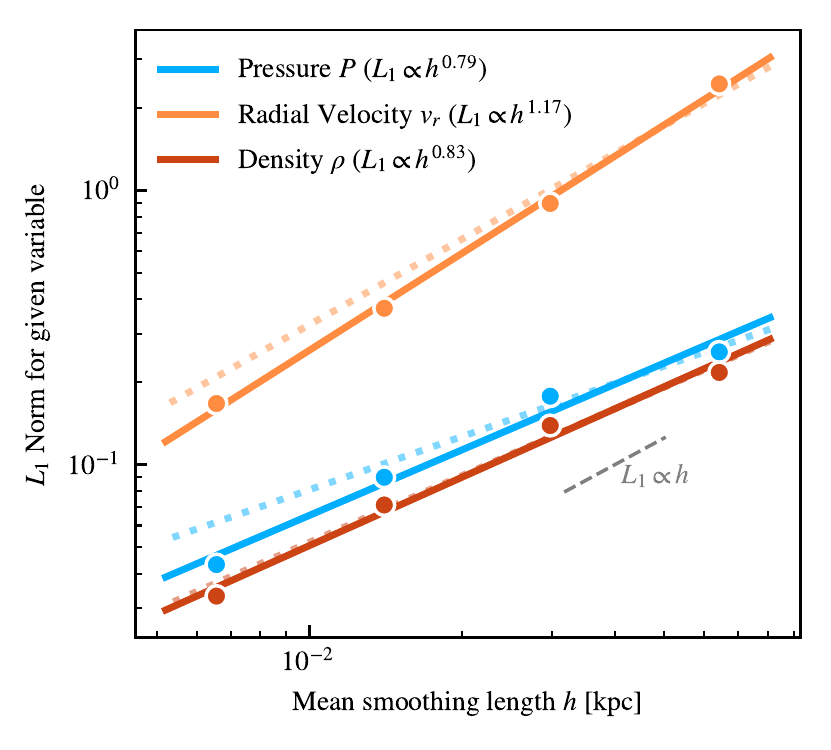}
    \vspace{-0.2cm}
    \caption{$L_1$ convergence for various gas properties for the Evrard
    collapse sphere at $t=0.8$. The region considered for convergence
    here is the purple band shown in Fig. \ref{fig:evrard}. The \sphenix{}
    scheme is shown with the points and linear fits in solid, and the same
    scheme is shown with artificial conduction turned off as dotted lines.
    Artificial conduction significantly improves convergence here as it
    helps stabilise the thermal properties of the initially randomly
    placed particles.}
    \label{fig:evrardconvergence}
\end{figure}

\subsection{Evrard Collapse}

The Evrard collapse \citep{Evrard1988} test takes a large sphere of
self-gravitating gas, at low energy and density, that collapses in on itself,
causing an outward moving accretion shock. This test is of particular interest for
cosmological and astrophysical applications as it allows for the inspection
of the coupling between the gravity and hydrodynamics solver.

\subsubsection{Initial Conditions}

Gas particles are set up in a sphere with an adiabatic index of $\gamma=5/3$,
sphere mass $M=1$, sphere radius $R=1$, initial density profile $\rho(r) = 1
/ 2\pi r$, and in a very cold state with $u = 0.05$, with the gravitational
constant $G=1$. These initial conditions are created in a box of size 100,
ensuring that effects from the periodic boundary are negligible.
Unfortunately, due to the non-uniform density profile, it is considerably
more challenging to provide relaxed initial conditions (or use a glass file).
Here, positions are simply drawn randomly to produce the required density
profile.

The Evrard collapse was performed at four resolution levels, with total
particle numbers in the sphere being $10^4$, $10^5$, $10^6$, and
$10^7$. The gravitational softening was fixed at $0.001$ for the $10^6$
resolution level, and this was scaled with $m^{-1/3}$ with $m$ the particle
mass for the other resolution levels. The simulations were performed once
with artificial conduction enabled (the full \sphenix{} scheme), and once
with it disabled.

\subsubsection{Results}

The highest resolution result ($10^7$ particles) with the full \sphenix{}
scheme is shown in Fig. \ref{fig:evrard}. This is compared against a high
resolution grid code\footnote{ {\tt HydroCode1D}, see
\url{https://github.com/bwvdnbro/HydroCode1D} and the \swift{} repository for
more details.} simulation performed in 1D, and here \sphenix{} shows an
excellent match to the reference solution. The shock at around $r=10^{-1}$ is
sharply resolved in all variables, and the density and velocity profiles show
excellent agreement. In the centre of the sphere, there is a slight deviation
from the reference solution for the internal energy and density (balanced to
accurately capture the pressure in this region) that remains even in the
simulation performed without artificial conduction (omitted for brevity, as
the simulation without conduction shows similar results to the simulation
with conduction, with the exception of the conduction reducing scatter in the
internal energy profile). This is believed to be an artefact of the initial
conditions, however it was not remedied by performing simulations at higher
resolutions.

The convergence properties of the Evrard sphere are shown in Fig.
\ref{fig:evrardconvergence}. The velocity profile shows a particularly
excellent result, with greater than linear convergence demonstrated. The
thermodynamic properties show roughly linear convergence. Of particular note
is the difference between the convergence properties of the simulations with
and without artificial conduction; those with this feature of \sphenix{}
enabled converge at a more rapid rate. This is primarily due to the
stabilising effect of the conduction on the internal energy profile. As the
particles are initially placed randomly, there is some scatter in the local
density field at all radii. This is quickly removed by adiabatic expansion in
favour of scatter in the internal energy profile, which can be stabilised by
the artificial conduction.

\begin{figure*}
    \centering
    \includegraphics{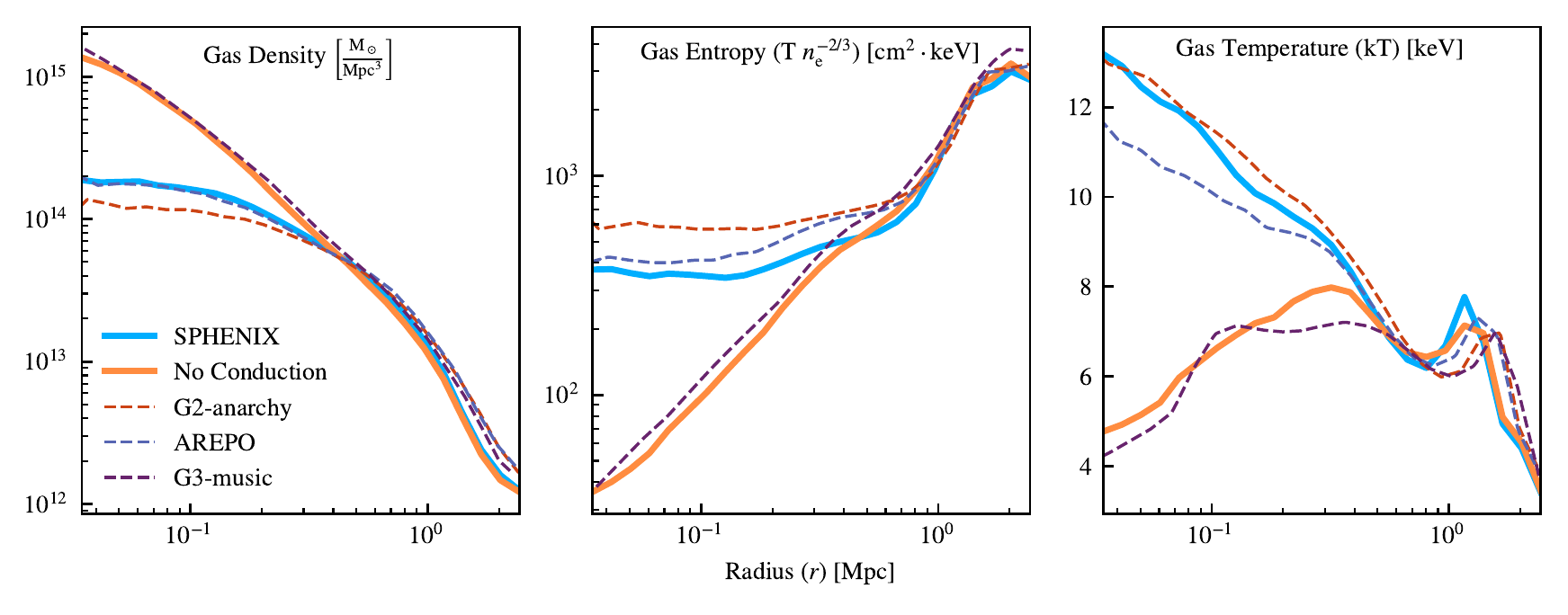}
    \vspace{-0.2cm}
    \caption{Thermodynamics profiles for the nIFTy cluster at $z=0$ with five
    codes and schemes. The solid lines show those simulated with \swift{},
    with the blue line showing the full \sphenix{} scheme, and the orange
    line showing \sphenix{} without artificial conduction
    . The dashed lines were extracted directly from
    \citet{Sembolini2016} and show a modern Pressure-Entropy scheme
    \citep[G2-anarchy;][appendix A]{Schaye2015}, a moving mesh finite volume
    scheme \citep[AREPO;][]{Springel2010}, and a traditional SPH scheme
    \citep[G3-music;][]{Springel2005}.}
    \label{fig:nifty}
\end{figure*}

\subsection{nIFTy Cluster}
\label{sec:niftycluster}

\begin{figure}
    \centering
    \includegraphics{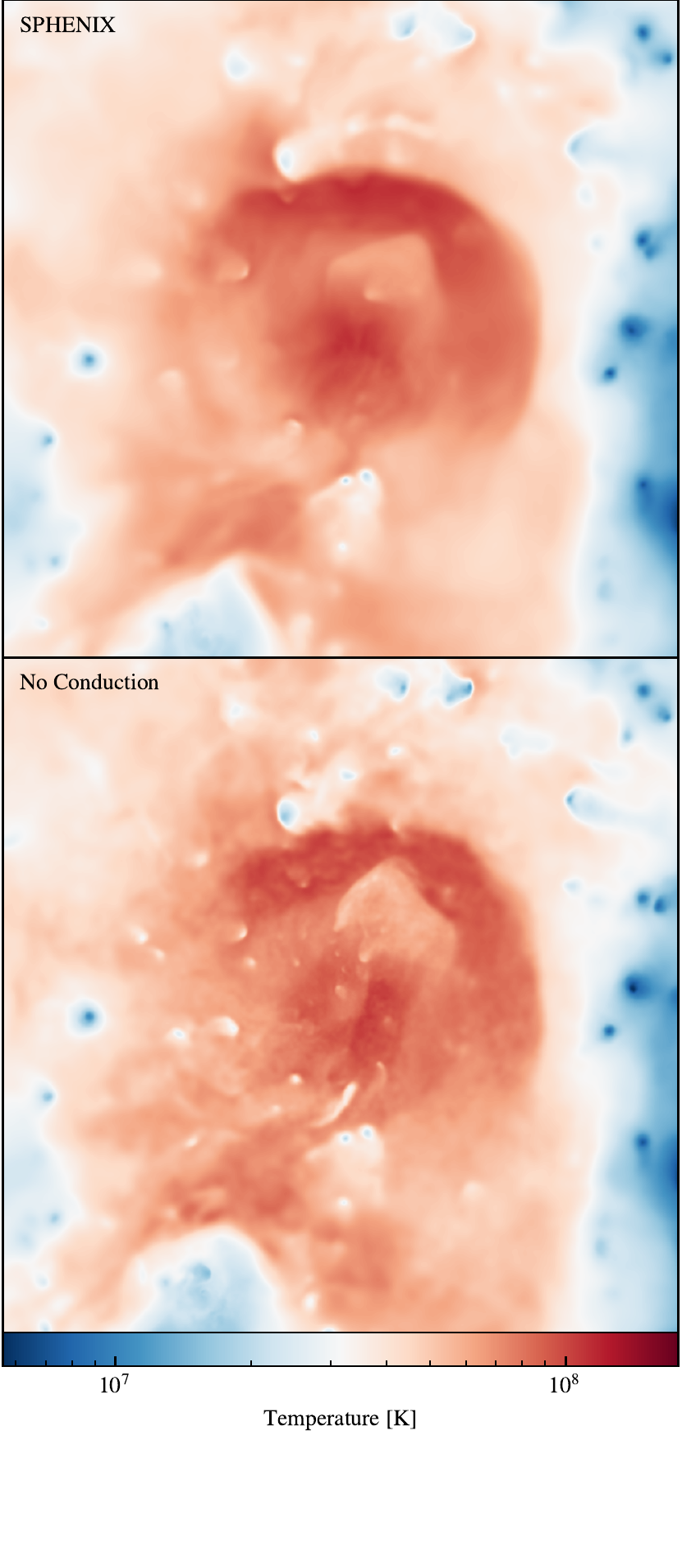}
    \vspace{-1.5cm}
    \caption{Image of the nIFTY cluster, as a projected mass-weighted
    temperature map, shown for the \sphenix{} scheme with (top) and without
    artificial conduction enabled (bottom). The image shows a 5 Mpc wide
    view, centred on the most bound particle in the halo.}
    \label{fig:niftyimage}
\end{figure}

The nIFTy cluster comparison project, \citet{Sembolini2016}, uses a
(non-radiative, cosmological) cluster-zoom simulation to evaluate the
efficacy of various hydrodynamics and gravity solvers. The original paper
compared various types of schemes, from traditional SPH
\citep[Gadget,][]{Springel2005} to a finite volume adaptive mesh refinement
scheme \citep[RAMSES,][]{Teyssier2002}. The true answer for this simulation is
unknown, but it is a useful case to study the different characteristics
of various hydrodynamics solvers. 

In Fig. \ref{fig:nifty} the \sphenix{} scheme is shown with and without
artificial conduction against three reference schemes from
\citet{Sembolini2016}. Here, the centre the cluster was found using the
VELOCIraptor \citep{Elahi2019} friends-of-friends halo finder, and the
particle with the minimum gravitational potential was used as the reference
point.

The gas density profile was created using 25 equally log-spaced radial bins,
with the density calculated as the sum of the mass within a shell divided by
the shell volume. \sphenix{} scheme shows a similar low-density core as
AREPO, with the no conduction scheme resulting in a cored density
profile similar to the traditional SPH scheme from \citet{Sembolini2016}.

The central panel of Fig. \ref{fig:nifty} shows the `entropy' profile of the
cluster; this is calculated as $T n_{\rm e}^{-2/3}$ with $n_{\rm e}$ the
electron density (assuming primordial gas, this is $n_{\rm e} = 0.875\rho /
m_{\rm H}$ with $m_{\rm H}$ the mass of a hydrogen atom) and $T$ the gas
temperature. Each was calculated individually in the same equally log-spaced
bins as the density profile, with the temperature calculated as the
mass-weighted temperature within that shell. The rightmost panel shows this
mass-weighted temperature profile, with \sphenix{} showing slightly higher
temperatures in the central region than AREPO, matching G2-anarchy instead.
This high-temperature central region, along with a low-density centre, lead
to the `cored' (i.e. flat, with high values of entropy, at small radii)
entropy profile for \sphenix{}.

The cored central entropy profile with \sphenix{} is attained primarily due
to the artificial conduction scheme and is not due to the other improvements
over the traditional SPH base scheme (including for example the artificial
viscosity implementation). We note again that there was no attempt to
calibrate the artificial conduction scheme to attain this result on the nIFTy
cluster, and any and all parameter choices were made solely based on the Sod
shock tube in \S \ref{sec:sodshock}.

In Fig. \ref{fig:niftyimage}, a projected mass-weighted temperature image of
the cluster is shown. The image demonstrates how the artificial conduction
present in the \sphenix{} scheme promotes phase mixing, resulting in the
cored entropy profile demonstrated in Fig. \ref{fig:nifty}.

The temperature distribution in the SPH simulation without conduction appears
noisier, due to particles with drastically different phases being present
within the same kernel. This shows how artificial conduction can lead to
sharper shock capture as the particle distribution is less susceptible to
this noise, enabling a cleaner energy transition between the pre- and
post-shock region.
\section{Conclusions}

We have presented the \sphenix{} SPH scheme and its performance on seven
hydrodynamics tests. The scheme has been demonstrated to show convergent
(with resolution) behaviour on all these tests. In summary:
\begin{itemize}
    \item \sphenix{} is an SPH scheme that uses Density-Energy
          SPH as a base, with added artificial viscosity for
          shock capturing and artificial conduction to reduce
          errors at contact discontinuities and to promote phase
          mixing.
    \item A novel artificial conduction limiter allows \sphenix{}
          to be used with energy injection feedback schemes (such
          as those used in \eagle{}) by reducing conduction across
          shocks and other regions where the artificial viscosity
          is activated.
    \item The artificial viscosity and conduction scheme coefficients
          were determined by ensuring good performance on the Sod
          Shock tube test, and remain fixed for all other tests.
    \item The modified Inviscid SPH \citep{Cullen2010} scheme captures
          strong shocks well, ensuring energy conservation, as
          shown by the Sedov-Taylor blastwave test, but the smooth
          nature of SPH prevents rapid convergence with resolution.
    \item The use of the \citet{Balsara1989} switch in \sphenix{}
          was shown to be adequate to ensure that the Gresho-Chan
          vortex is stable. Convergence on this test
          was shown to be faster than in \citet{Cullen2010}.
    \item The artificial conduction scheme was shown to work adequately
          to capture thermal instabilities in both the Kelvin-Helmholtz
          and Blob tests, with contact discontinuities well preserved
          when required.
    \item \sphenix{} performed well on both the Evrard collapse and nIFTY
          cluster problems, showing that it can couple to the
          FMM gravity solver in \swift{} and that the artificial
          conduction scheme can allow for entropy cores in clusters.
    \item \sphenix{} is implemented in the \swift{} code and is
          available fully open source to the community.
\end{itemize}

\sphenix{} hence achieves its design goals; the Lagrangian nature of the
scheme allows for excellent coupling with gravity; the artificial conduction
limiter allows the injection of energy as in the \eagle{} sub-grid physics
model; and the low cost-per-particle and lack of matrices carried on a
particle-by-particle basis provide for a very limited computational cost
\citep[see ][ for a comparison of computational costs between a scheme like
\sphenix{} and the GIZMO-like schemes also present in \swift{}]{Borrow2019}.
\section{Acknowledgements}
\label{sec:acknowledgements}

The authors thank Folkert Nobels for providing initial conditions for Appendix
\ref{app:nfw}, and the anonymous referee for their comments that improved the
paper.
JB is supported by STFC studentship ST/R504725/1.
MS is supported by the Netherlands Organisation for Scientific Research (NWO) through VENI grant 639.041.749.
RGB is supported by the Science and Technology Facilities Council ST/P000541/1.
This work used the DiRAC@Durham facility managed by the Institute for
Computational Cosmology on behalf of the STFC DiRAC HPC Facility
(www.dirac.ac.uk). The equipment was funded by BEIS capital funding
via STFC capital grants ST/K00042X/1, ST/P002293/1, ST/R002371/1 and
ST/S002502/1, Durham University and STFC operations grant
ST/R000832/1. DiRAC is part of the National e-Infrastructure.

\subsection{Software Citations}

This paper made use of the following software packages:
\begin{itemize}
    \item {\sc Swift} \citep{Schaller2018}
    \item {\tt python} \citep{VanRossum1995}, with the following libraries
    \begin{itemize}
    	\item {\tt numpy} \citep{Harris2020}
    	\item {\tt scipy} \citep{SciPy1.0Contributors2020}
    	\item {\tt numba} \citep{Lam2015}
    	\item {\tt matplotlib} \citep{Hunter2007}
    	\item {\tt swiftsimio} \citep{Borrow2020a}
    \end{itemize}
\end{itemize}

\section{Data Availability}
\label{sec:dataavail}

All code and initial conditions used to generate the simulations is open
source as part of \swift{} version 0.9.0 \citep{Schaller2018}\footnote{
Available from \url{http://swift.dur.ac.uk},
\url{https://gitlab.cosma.dur.ac.uk/swift/swiftsim}, and
\url{https://github.com/swiftsim/swiftsim}. }. As the simulations presented
in this paper are small, test, simulations that can easily be repeated, the
data is not made immediately available.

\bibliographystyle{mnras}
\bibliography{newbib}

\appendix
\section{Particle Costs}
\label{app:particlecost}

\begin{figure}
    \centering
    \includegraphics{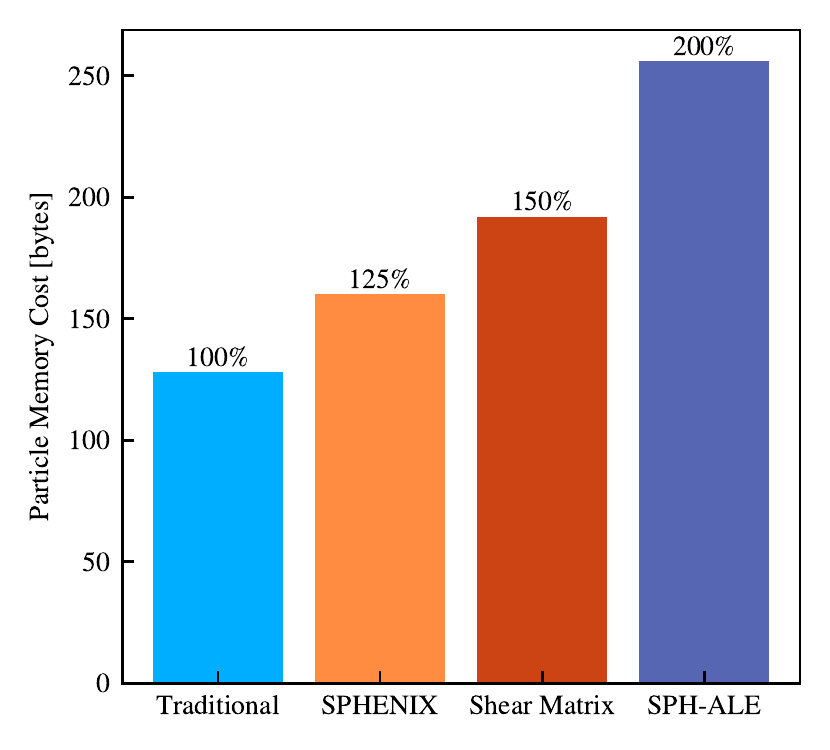}
    \caption{Cost per particle (in bytes) for four different hydrodynamics
    models (see text for details). Percentages are give relative to the
    Traditional (similar to Gadget-2, with no artificial conduction and
    fixed artificial viscosity coefficients) model.}
    \label{fig:costs}
\end{figure}

Different SPH models require different information stored per particle.
Compared to a basic, `Traditional' SPH model, \sphenix{} requires a 
small amount of extra data to store things like the particle-carried artificial conduction
and viscosity coefficients. The amount of data required increases for
more complex models, such as those making use of the full shear tensor,
like \citet{Wadsley2017}, or additional corrections, like \citet{Rosswog2020}.
SPH models using an ALE (Arbitrary Lagrangian-Eulerian) framework \citep[see][]{Vila1999}
require even more information as the particles carry flux information for use
in the Riemann solver.

The amount of data required to store a single element in memory is of upmost
importance when considering the speed at which a simulation will run. SPH codes,
and \swift{} in particular, are bound by the memory bandwidth available, rather
than the costs associated with direct computation. This means any increase in
particle cost corresponds to a linear increase in the required computing time for
simulation; this is why keeping the particles lean is a key requirement of the
\sphenix{} model. Additionally, in large simulations performed over many nodes,
the bandwidth of the interconnect can further become a limitation and hence
keeping the memory cost of particles low is again beneficial.

In Fig. \ref{fig:costs} we show the memory cost of four models: Traditional SPH
\citep[similar to to the one implemented in Gadget-2;][]{Springel2005},
\sphenix{}, a model with the full shear matrix, and a SPH-ALE model similar to
GIZMO-MFM \citep{Hopkins2015}, all implemented in the \swift{} framework. We see
that \sphenix{} only represents a 25\% increase in memory cost per particle for
significant improvement over the traditional model.
\section{Conduction Speed}
\label{app:conduction}

\begin{figure*}
    \centering
    \includegraphics{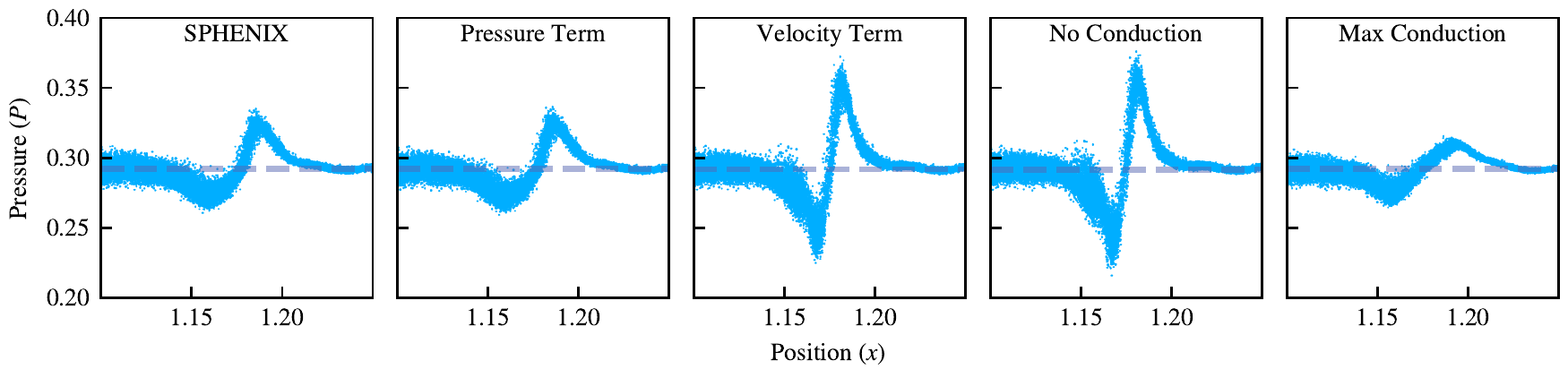}
    \caption{The pressure contact discontinuity in the Sod Shock (Fig.
    \ref{fig:sodshock}) at a resolution of $32^3$ and $64^3$, but using glass
    files instead of the BCC lattices (this leads to significantly increased
    particle disorder, but more evenly distributes particles in the $x$ direction
    enabling this figure to be clearer). Here we show a zoomed-in representation
    of all particles (blue points) against the analytical solution (purple dashed
    line). Each sub-figure shows the simulation at the same time $t=0.2$, but with
    different forms for the conduction velocity (see text for details).}
    \label{fig:condspeed}
\end{figure*}

The conduction speed (Eqn. \ref{eqn:condspeed}) in \sphenix{} was primarily
selected for numerical reasons. In a density based scheme, it is common to see
significant errors around contact discontinuities where there are large changes
in density and internal energy simultaneously to produce a uniform pressure
field. In Fig. \ref{fig:sodshock} we demonstrated the performance of the
\sphenix{} scheme on one of these discontinuities, present in the Sod Shock.

In Fig. \ref{fig:condspeed} we zoom in on the contact discontinuity, this time
using glass files for the base initial conditions (of resolution $32^3$ and
$64^3$), allowing for a more even distribution of particles along the horizontal
axis. We use five different models,
\begin{itemize}
    \item \sphenix{}, the full \sphenix{} model using the conduction speed from Eqn. \ref{eqn:condspeed}.
    \item Pressure Term, which uses only the pressure-based term from Eqn. \ref{eqn:condspeed}.
    \item Velocity Term, which only uses the velocity-based term from Eqn. \ref{eqn:condspeed}.
    \item No Conduction, which sets the conduction speed to zero.
    \item Max Conduction, which sets $\alpha_{\rm D}$ to unity everywhere, and uses the conduction speed from Eqn. \ref{eqn:condspeed}.
\end{itemize}
The first thing to note here is that the pressure term provides the vast majority
of the conduction speed, highlighting the importance of this form of bulk conduction
in \sphenix{} and other models employing a density based equation of motion. Importantly,
the conduction allows for the `pressure blip' to be reduced to a level where there is no
longer a discontinuity in pressure (i.e. there is a smooth gradient with $x$).
Although the velocity term is able to marginally reduce the size of the blip
relative to the case without conduction, it is unable to fully stabilise the
solution alone. Pressure blips can lead to large pressure differences between
individual particles, then leading to the generation of a divergent flow around
the point where the contact discontinuity resides. This is the primary motivation
for the inclusion of the velocity divergence-based term in the conduction
speed. Along with the conduction limiter (see Eqn. \ref{eqn:diffK} for the source
term), if there is a large discontinuity in internal energy that is generating
a divergent flow (and not one that is expected to do so, such as a shock), the
velocity-dependent term can correct for this and smooth out the internal energy
until the source of divergence disappears.

\subsection{Alternative Conduction Speeds}

\begin{figure*}
    \centering
    \includegraphics{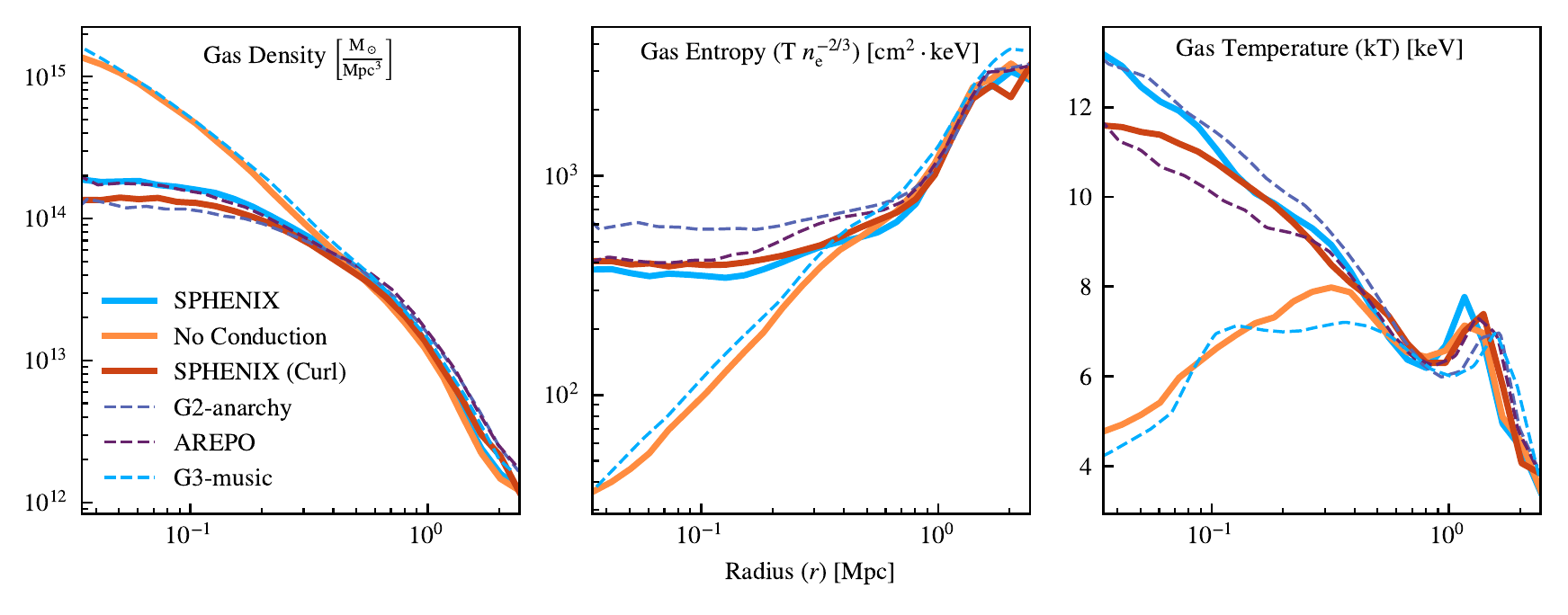}
    \caption{Reproduction of Fig. \ref{fig:nifty} but including the line
    (red) for the version of \sphenix{} performed with an explicit shear component
    in the conduction speed. We see no qualitative differences between the two models,
    with them both providing an entropy core at a similar level.}
    \label{fig:niftycomparespeed}
\end{figure*}

The \sphenix{} conduction speed (Eqn. \ref{eqn:condspeed}) contains two components: one
based on pressure differences and one based on a velocity component. In \sphenix{},
as in a number of other models, this velocity component really encodes compression
or expansion along the axis between particles.

The motivation for some alternative schemes \citep[e.g. those presented in][]{Wadsley2008,
Wadsley2017} is shear between particles. To test if we see significant differences
in our tests, we formulate a new conduction speed,
\begin{equation}
    v_{D, ij} = \frac{\alpha_{D, ij}}{2}
    \left(
        \frac{|\vec{v}_{ij} \times \vec{x}_{ij}|}{|\vec{x}_{ij}|} + 
        \sqrt{2\frac{|P_i - P_j|}{\hat{\rho}_j + \hat{\rho}_j}}
    \right).
    \label{eqn:altcondspeed}
\end{equation}
that focuses on capturing the shear component of the velocity between two particles.

\begin{figure}
    \centering
    \includegraphics{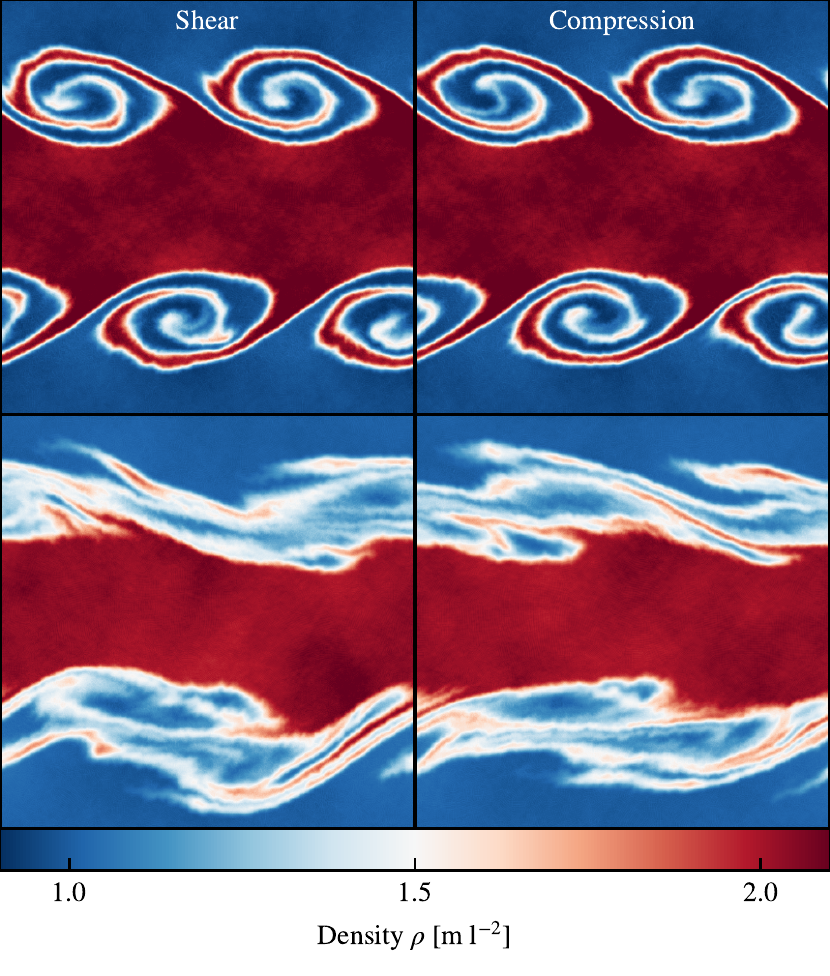}
    \caption{Kelvin-Helmholtz test with a density contrast of $\rho_{\rm C}=2$
    as in Fig. \ref{fig:kelvinhelmholtz}, shown at $t=2\tau_{\rm KH}$ (top) and
    $t=4\tau_{\rm KH}$ (bottom). We show on the left the simulation with the
    shear-based conduction speed, and again the compression-based speed on the
    right. No significant qualitative differences are seen between the two
    models.}
    \label{fig:khcomparespeed}
\end{figure}

We again test this new formulation on some of our example problems. First,
the nIFTy cluster, presented in Fig. \ref{fig:niftycomparespeed}, shows little
difference between the two formulations, with both providing a solution similar
to other modern SPH schemes and grid codes.

The Kelvin-Helmholtz test again shows little difference (Fig.
\ref{fig:khcomparespeed}), although there is a slightly increased growth rate of
the perturbation at late times for the shear formulation.

We find no discernible difference between the two formulations on the blob test,
as this is mainly limited by the choice of Density-SPH as the base equation of
motion to correctly capture the initial break up of the blob from the centre outwards.
\section{Maintenance of Hydrostatic Balance}
\label{app:nfw}

\begin{figure*}
    \centering
    \includegraphics{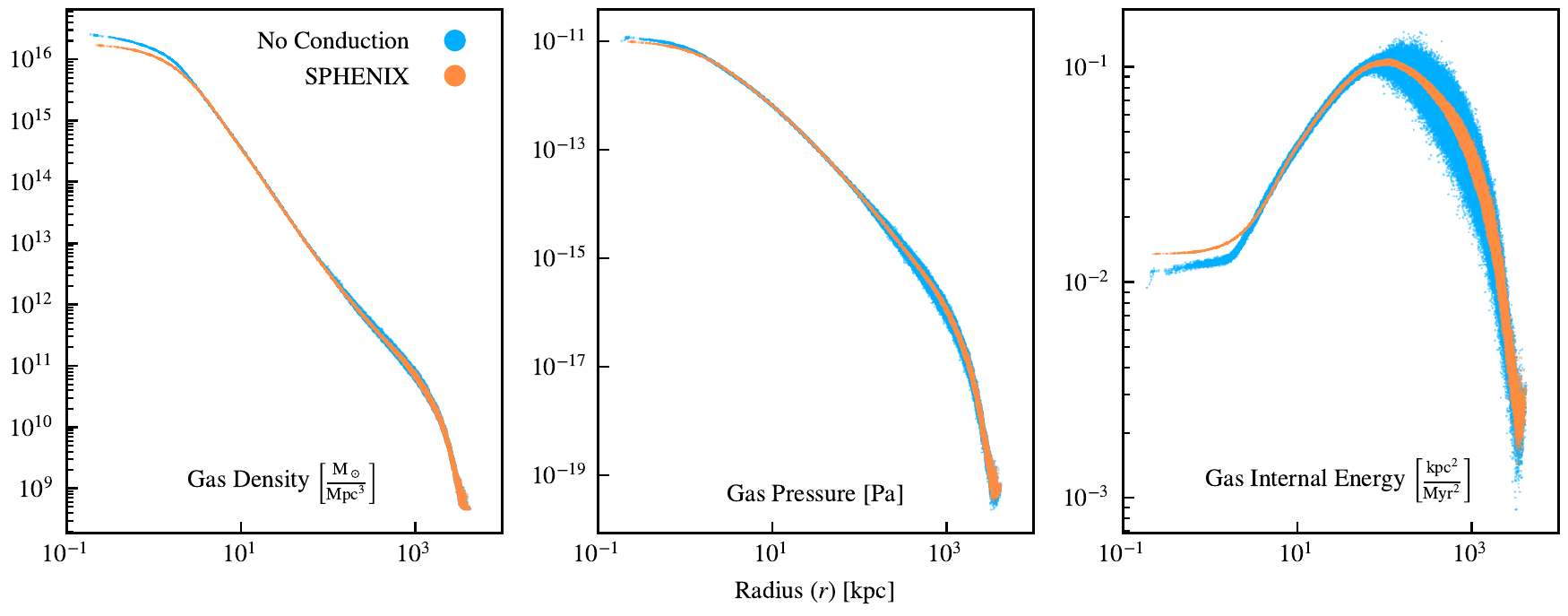}
    \caption{Profiles of the idealised NFW halo (of mass $\approx 10^{13}$
    M$_\odot$, at a gas particle mass resolution of $10^5$ M$_\odot$) at $t=5$
    Gyr after the initial state. Blue points show every particle in the
    simulation without artificial conduction enabled, with orange showing the
    simulation with conduction enabled. Here the conduction can allow for a
    reduction in the scatter in internal energy without leading to significant
    conduction into the centre. The offset seen in the centre of about a factor
    of 1.5x originates from the smoothing of the kink around $\approx 0.7$ kpc
    during the initial settling of the halo, and remained stable from that point
    at around $\approx t=0.5$ Gyr until the end of the simulation.}
    \label{fig:nfw}
\end{figure*}

The form of the conduction speed used in \sphenix{} based on pressure
differences (Eqn. \ref{eqn:condspeed}) has been conjectured to not allow for the
maintenance of a pressure gradient against some external body force \citep[for
example a halo in hydrostatic equilibrium;][]{Sembolini2016}. The main concern
here is that the pressure difference form of the conduction speed may allow
thermal energy to travel down into the gravitational potential, heating the
central regions of the halo.  As \sphenix{} uses an additional limiter (Eqn.
\ref{eqn:diffK} for the source term) that only activates conduction in regions
where the internal energy gradient cannot be represented by SPH anyway, this may
be less of a concern.  Additionally, there will be no conduction across
accretion shocks due to the limiter in Eqn. \ref{eqn:condshocklimiter}.

In Fig. \ref{fig:nfw} we show an idealised simulation of an adiabatic halo with
an NFW \citep{Navarro1996} dark matter density profile, and gas in hydrostatic
equilibrium.  The halo uses a fixed NFW potential in the background, with a mass
of $10^{13}$ M$_\odot$, concentration $7.2$, and a stellar bulge fraction of
1\%. The halo has a gas mass of $1.7 \times 10^{12}$ M$_\odot$, resolved by
1067689 particles with varying mass from $10^5$ to $1.7 \times 10^{12}$
M$_\odot$ with the highest resolution in the centre.

The gas in the halo is set up to be isothermal, following \citep{Stern2019},
\begin{equation}
    \frac{\mathrm{d} \, \ln P}{\mathrm{d} \, \ln r} = 
    - \gamma \frac{v_c^2}{c_s^2}
\end{equation}
where $v_c$ is the circular velocity of the halo. The condition used to set
the initial temperatures is $v_c = c_s$, and to get the correct normalisation
for pressure and density the gas fraction at $R_{500, {\rm crit}}$ is used
following \citet{Debackere2020}.

Fig. \ref{fig:nfw} shows that there is little difference between the result
with conduction, and without. There is a small offset in the centre where the
simulation with conduction has a slightly higher energy and slightly lower
density, giving a very small overall offset in pressure. This figure is shown
at $t=5$ Gyr, much longer than any realistic cluster of a similar mass would 
go without accretion or some other external force perturbing the pressure
profile anyway. Finally, the conduction allows the noisy internal energy
distribution (and additionally density distribution) to be normalised over time
thanks to the inclusion of the pressure differencing term.
\section{Sedov Blast}
\label{app:sedovblast}

\begin{figure*}
    \centering
    \includegraphics{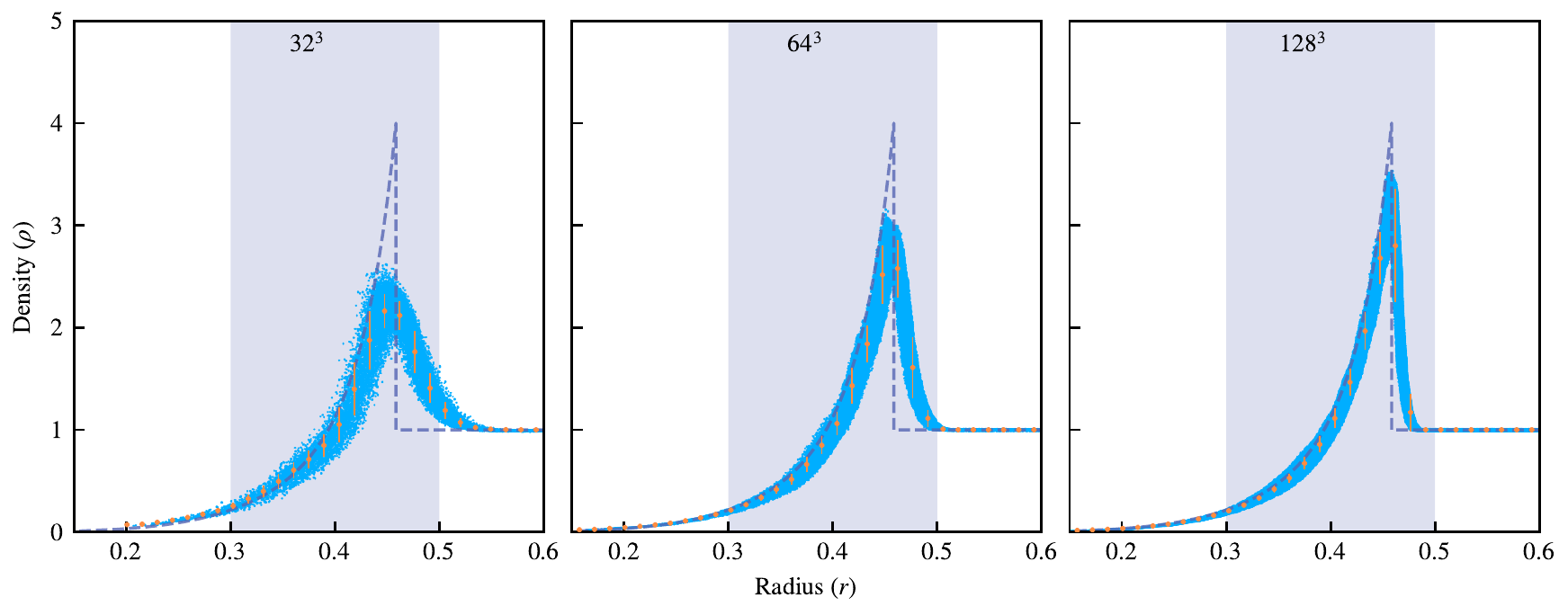}
    \vspace{-0.2cm}
    \caption{The density profile of the Sedov blasts initially presented
    in Fig. \ref{fig:sedovconvergence}. The blue points show the positions
    of every particle in the volume, the purple dashed line the analytical prediction,
    and the orange points binned means with error bars showing one standard
    deviation. The shaded band is the region over which the convergence
    properties were measured. The text at the top notes the total number of
    particles in each volume.}
    \label{fig:sedovmultiple}
\end{figure*}

In Fig. \ref{fig:sedovconvergence} we presented the convergence properties of
the Sedov blast with the \sphenix{} scheme. The scheme only demonstrated
convergence as $L_1(v) \propto h^{\sim 0.5}$, which is much slower than the
expected convergence rate of $L_1 \propto h^1$ for shock fronts in SPH (that
is demonstrated and exceeded in the Noh problem in Fig. \ref{fig:nohconv}).
This is, however, simply an artefact of the way that the convergence is
measured.

In Fig. \ref{fig:sedovmultiple} we show the actual density profiles of
the shock front, by resolution (increasing as the subfigures go to the
right). Note here that the width of the shock front (from the particle
distribution to the right of the vertical line to the vertical line
in the analytical solution) does converge at the expected rate of 
$L_1 \propto 1 / n^{1/3} \propto h$ with $n$ the number of particles in the volume
(in 3D).

The Sedov blast, unlike the Noh problem and Sod tubes, does not aim to
reproduce a simple step function in density and velocity, but also a complex,
expanding, post-shock region. The $L_1$ convergence is measured `vertically'
in this figure, but it is clear here that the vertical deviation from the
analytical solution is not representative of the `error' in the properties of
a given particle, or in the width of the shock front. Small deviations in 
the position of the given particle could result in changes of orders of
magnitude in the value of the $L_1$ norm measured for it.

Because of this, and because we have demonstrated in other sections that
\sphenix{} is able to converge on shock problems at faster than first order,
we believe the slow convergence on the Sedov problem to be of little
importance in practical applications of the scheme.
\section{Conduction in the Noh Problem}
\label{app:nohcond}

\begin{figure*}
    \centering
    \includegraphics{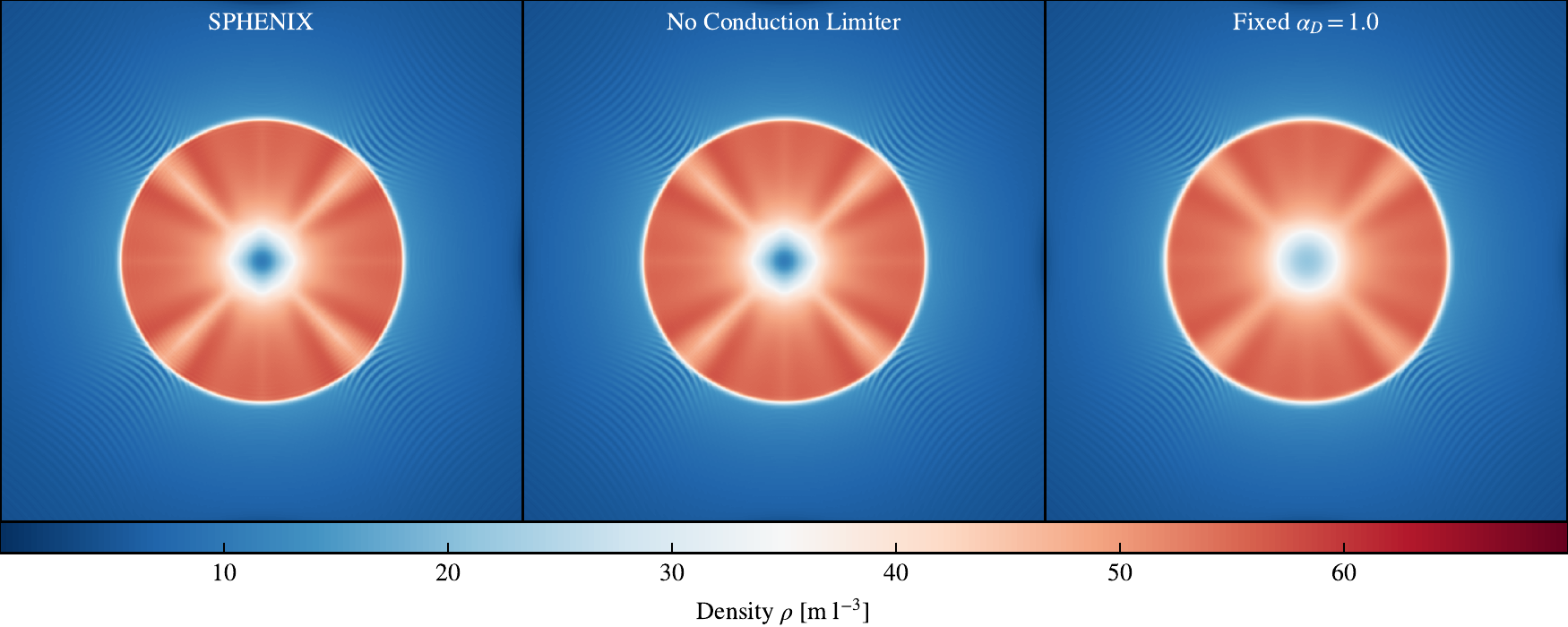}
    \caption{Density slice through the centre of the Noh problem
    (analogue of Fig. \ref{fig:nohimg}) shown for three different
    artificial conduction schemes (see text). The colour bar is
    shared between all, and they all use the same, $256^3$,
    initial condition, and are also all shown at $t=0.6$. The
    case with the fixed, high, conduction coefficient (right)
    shows the smallest deviation in density in the centre,
    as the conduction can treat the wall heating present in
    this test.}
    \label{fig:nohcomparediffusion}
\end{figure*}

In \S \ref{sec:noh} we presented the Noh problem, and showed that the
\sphenix{} scheme (like other SPH schemes in general) struggles to capture
the high density in the central region due to so-called `wall heating'.

The \sphenix{} scheme includes a switch to reduce artificial conduction
in viscous flows. This is, as presented in \S \ref{sec:conductionlimiter},
to allow for the capturing of energetic feedback events. It does, however,
lead to a minor downside; the stabilising effect of the conduction in
these shocks is almost completely removed. Usually, the artificial conduction
lowers the dispersion in local internal energy values, and hence pressures,
allowing for a more regular particle distribution. 

In Fig. \ref{fig:nohcomparediffusion} we show three re-simulations of the Noh
problem (at $256^3$ resolution) with three separate schemes. The first, the
full \sphenix{} scheme, is simply a lower resolution version of Fig.
\ref{fig:nohimg}. The second, `No Conduction Limiter', is the \sphenix{}
scheme, but with Equation \ref{eqn:conductionlimiter} removed; i.e. the
particle-carried artificial conduction coefficient depends solely on the
local internal energy field (through $\nabla^2 u$ and Eqn. \ref{eqn:diffK}),
instead of also being mediated by the velocity divergence field. The final
case, `Fixed $\alpha_D=1.0$', shows the case where we remove all conduction
switches and use a fixed value for the conduction $\alpha_D$ of $1.0$. The
former two look broadly similar, suggesting that the post-shock region is not
significantly affected by the additional \sphenix{} conduction limiter.

The final panel, however, shows the benefits available to a hypothetical
scheme that can remove the artificial conduction switch; the central region
is able to hold a significantly higher density thanks to energy being
conducted out of this region, allowing the pressure to regularise. In
addition to the above, this case shows significantly weaker spurious density
features (recall that the post-shock, high-density, region should have a
uniform density) because these have been regularised by the conduction
scheme.

We present this both to show the drawbacks of the \sphenix{} artificial
conduction scheme, and to show the importance of demonstrating test problems
with the same switches that would be used in a production simulation.
\section{Blob Test}
\label{app:blob}

\begin{figure}
    \includegraphics{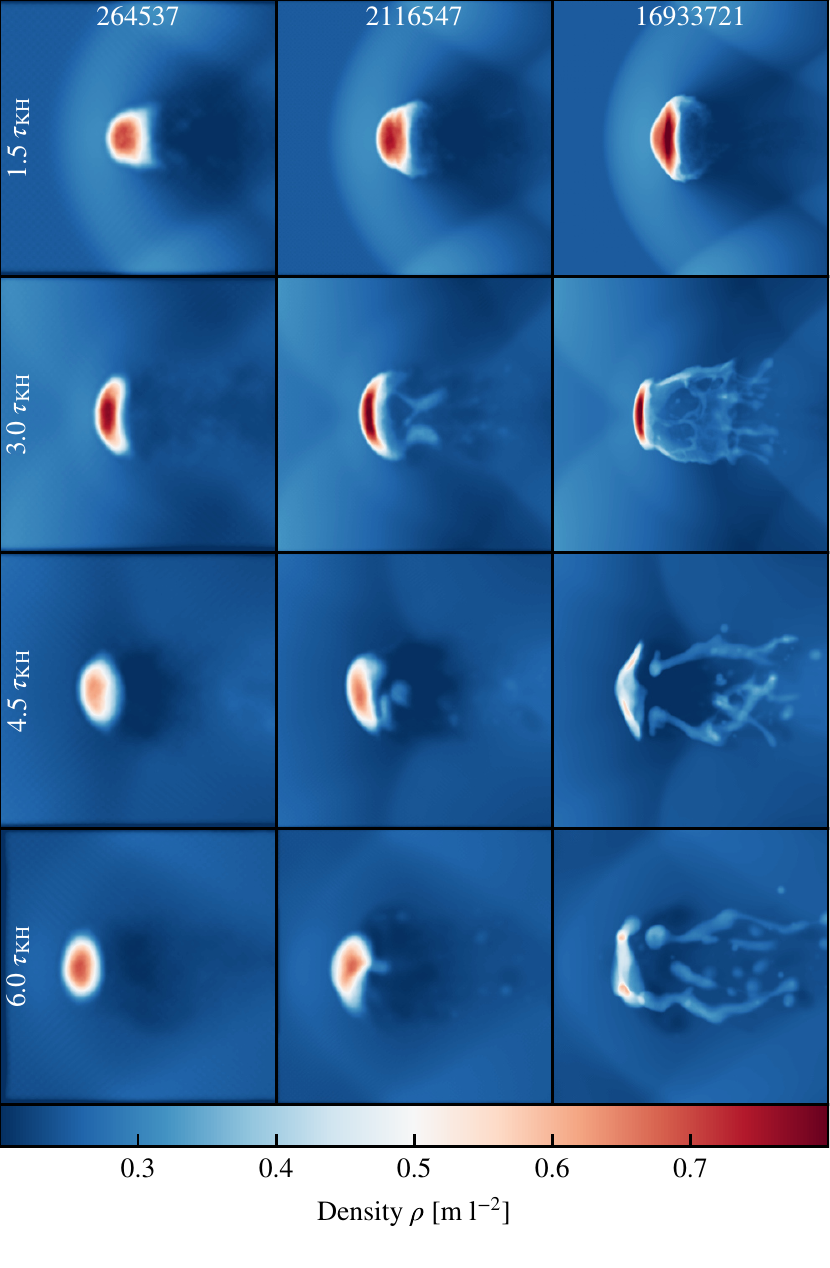}
    \vspace{-0.2cm}
    \caption{A repeat of Fig. \ref{fig:blob} but using a `traditional' SPH
    scheme without diffusive switches.}
    \label{fig:blob_minimal}
\end{figure}

\begin{figure}
    \includegraphics{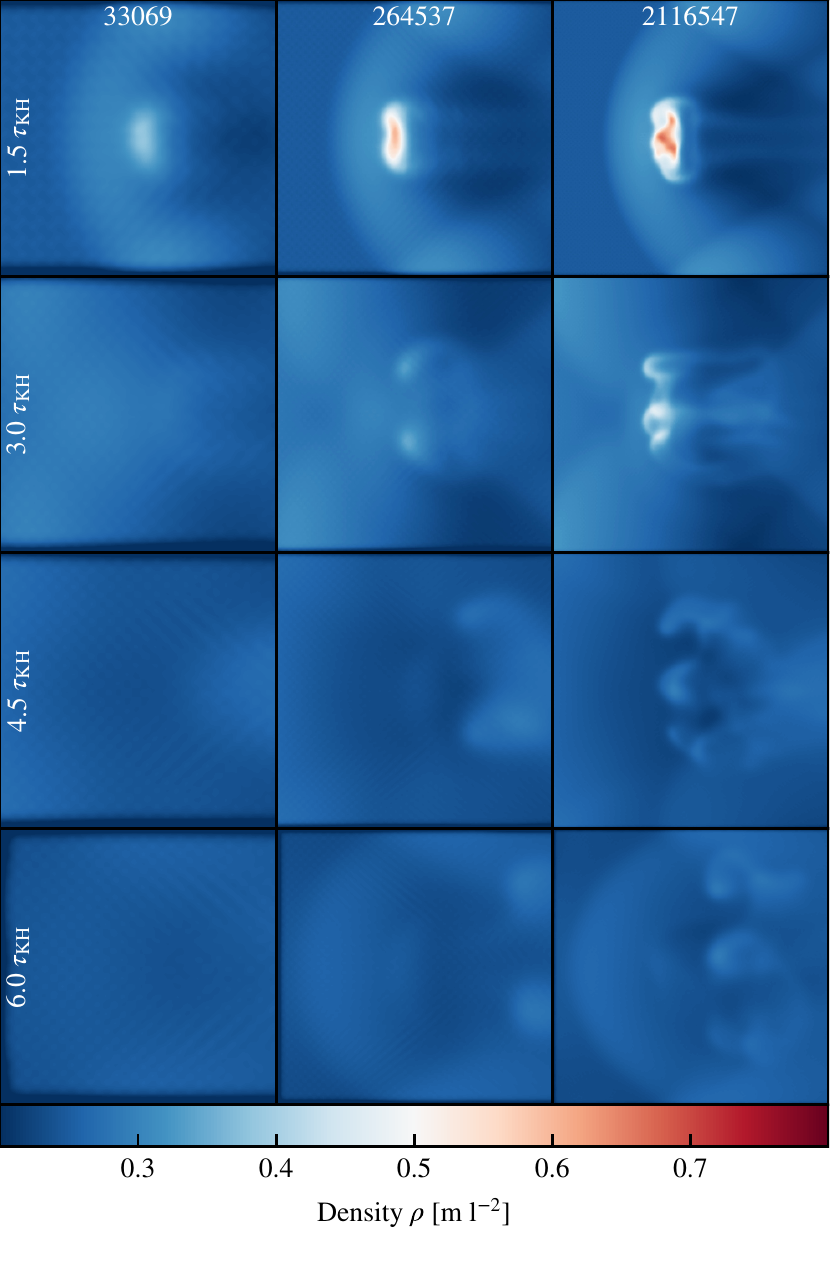}
    \vspace{-0.2cm}
    \caption{A repeat of Fig. \ref{fig:blob} but using an SPH-ALE
    scheme with a diffusive slope limiter. Note however that this is
    one step lower in resolution, due to the additional computational
    cost required to perform a simulation including a Riemann solver.}
    \label{fig:blob_gizmo}
\end{figure}

\begin{figure*}
    \includegraphics{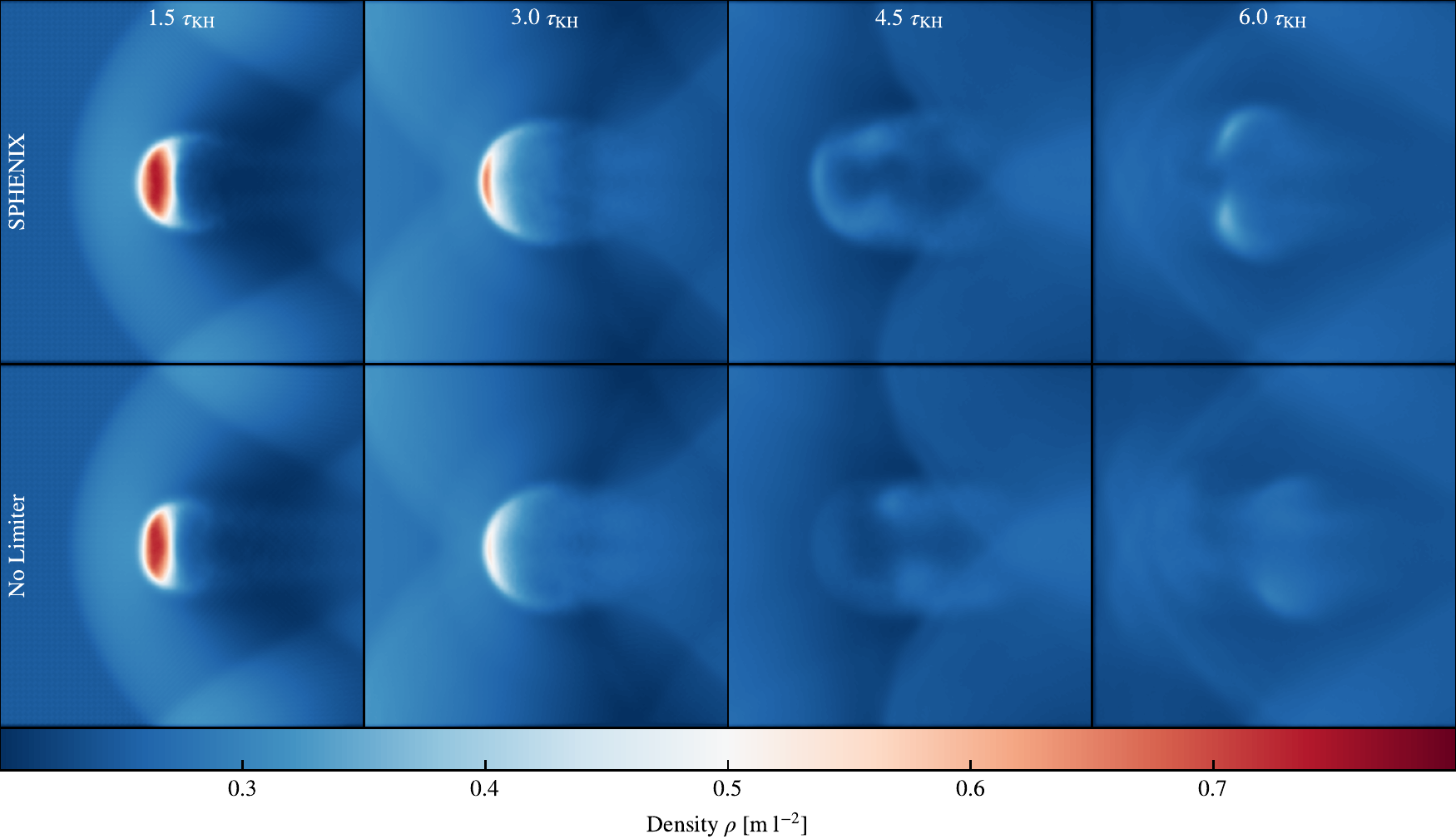}
    \caption{The evolution of a single blob (using the medium resolution,
    2116547 particle, initial conditions from Fig. \ref{fig:blob}), to
    illustrate the effect of turning off the conduction limiter (Eqn.
    \ref{eqn:conductionlimiter}; bottom row) in comparison to the full
    \sphenix{} scheme (top row). The limiter suppresses some of the initial
    mixing during the cloud crushing, but does not cause significant
    qualitative changes in the mixing of the cloud.}
    \label{fig:blob_limiter}
\end{figure*}

In Fig. \ref{fig:blob} we demonstrated the performance of the \sphenix{}
scheme on an example `blob' test. Here, we show how the same initial
conditions are evolved using two schemes: a `traditional SPH' scheme with
fixed artificial viscosity ($\alpha_V = 0.8$) and no artificial conduction
\citep[e.g.][]{Monaghan1992}\footnote{The {\tt minimal} scheme in \swift{}.},
and a SPH-ALE \citep{Vila1999} scheme similar to GIZMO-MFM\footnote{The {\tt
gizmo-mfm} scheme in \swift{} with a HLLC Riemann solver.}
\citep{Hopkins2015} with a diffusive slope limiter. This is in an effort to
demonstrate how the initial conditions are evolved with a minimally viable
non-diffusive scheme, through to what could be considered the most diffusive
viable scheme.

Fig. \ref{fig:blob_minimal} shows the result of the blob test with the
traditional SPH scheme. Here, as expected, there is a severe lack of mixing,
with the artificial surface tension holding the blob together even at the
highest resolutions. The lack of phase mixing also contributes to a lack
of overall mixing, with the stripped trails (shown most clearly at
$t=3\tau_{\rm KH}$) adiabatically expanding but crucially remaining
distinct from the hot background medium.

Fig. \ref{fig:blob_gizmo} shows the result of the blob test with the SPH-ALE
(GIZMO-MFM) scheme. This scheme is known to be highly diffusive (due to the
less conservative slope limiter employed in the \swift{} implementation).
This follows closely the results seen in e.g. \citet{Agertz2007} for
diffusive grid-based codes. Here, the blob is rapidly shattered, and then
dissolves quickly into the surrounding media, especially at the lowest
resolutions.

The \sphenix{} results in Fig. \ref{fig:blob} showed that the blob
mixed with the surrounding media, but at a less rapid rate than in
the SPH-ALE case. This is somewhat expected, given the trade-off
required in the artificial conduction switches (Eqn.
\ref{eqn:conductionlimiter}). We do note, however, that no analytical
solution exists for the blob test, and as such all of these comparisons
may only be made qualitatively.

In Fig. \ref{fig:blob_limiter} we examine the effect of removing the
conduction limiter from the \sphenix{} implementation (i.e. Eqn.
\ref{eqn:conductionlimiter} is removed, allowing $\alpha_D$ to vary
irrespective of the values of $\alpha_V$). We see that the inclusion of the
limiter does slightly reduce the rate of initial mixing within the blob, but
that the effect of the limiter is not particularly strong within this case.

\end{document}